\documentclass[notitlepage,nofootinbib,preprintnumbers,amssymb,superscriptaddress]{revtex4-1}

\usepackage{amsfonts,amssymb,amsmath,graphicx,color,bm,multirow}
\usepackage[utf8]{inputenc}
\definecolor{ultramarine}{rgb}{0.07, 0.04, 0.56}
\definecolor{cadmiumgreen}{rgb}{0.0, 0.42, 0.24}
\definecolor{indigo(dye)}{rgb}{0.0, 0.25, 0.42}

\newcommand{\Mpl}{M_{\mathrm{pl}}}
\newcommand{\vDM}{v_\mathrm{DM}}
\newcommand{\e}{\mathrm{e}}
\newcommand{\rhoDM}{\rho_{\mathrm{DM}}}
\newcommand{\Tobs}{T_{\mathrm{obs}}}
\newcommand{\cc}{\mathrm{c.c.}}
\newcommand{\omegak}{\omega_{\vec{k}}}
\newcommand{\veck}{\vec{k}}
\newcommand{\phik}{\phi_{\vec{k}}}
\newcommand{\thetak}{\theta_{\vec{k}}}
\newcommand{\Od}{\Omega_{\mathrm{d}}}
\newcommand{\Oy}{\Omega_{\mathrm{y}}}
\newcommand{\deltar}{\delta_{\mathrm{r}}}
\newcommand{\Ndet}{N_{\mathrm{det}}}
\newcommand{\fdet}{f_{\mathrm{det}}}

\newcommand{\rhoc}{\rho_{\mathrm{c}}}
\newcommand{\Nt}{N_{\mathrm{t}}}

\newcommand{\hth}{h_{\mathrm{th}}}

\begin{document}

\title{On the detectability of ultralight scalar field dark matter\\with gravitational-wave detectors}

\author{Soichiro Morisaki}
\affiliation{Research Center for the Early Universe (RESCEU), Graduate School of Science, The University of Tokyo, Tokyo 113-0033, Japan}
\affiliation{Department of Physics, Graduate School of Science, The University of Tokyo, Tokyo 113-0033, Japan}
\author{Teruaki Suyama}
\affiliation{Department of Physics, Tokyo Institute of Technology, 2-12-1 Ookayama, Meguro-ku, 
Tokyo 152-8551, Japan}

\begin{abstract}
An ultralight scalar field is one of the dark matter candidates.
If it couples with Standard Model particles, it oscillates mirrors in gravitational-wave detectors 
and generates detectable signals.
We study the spectra of the signals taking into account the motion of the detectors due to the Earth's rotation/the detectors' orbital motion around the Sun
and formulate a suitable data-analysis method to detect it.
We find that our method can improve the existing constraints given by fifth-force experiments on one of the scalar field's coupling constants by a factor of $\sim 30$, $\sim 100$ and $\sim 350$ for $m_\phi = 2 \times 10^{-17}~\mathrm{eV},~10^{-14}~\mathrm{eV}$ and $10^{-12}~\mathrm{eV}$ respectively, where $m_\phi$ is the scalar field's mass.
Our study demonstrates that
experiments with gravitational-wave detectors play a complementary role to that Equivalence Principle tests do.
\end{abstract}

\maketitle

\section{Introduction} \label{Introduction}

Although Weakly Interacting Massive Particles (WIMPs) are promising candidates of dark matter, null results from various experiments \cite{Aprile:2018dbl,Ackermann:2015zua,Sirunyan:2017hci,Aaboud:2016tnv} cast doubt on WIMPs.
Therefore, it is worth searching for other candidates.
Ultralight scalar field is one of the other dark matter candidates.
This type of dark matter can have mass down to $\sim 10^{-22}~\mathrm{eV}$ \cite{Hu:2000ke} and we especially consider a mass range of $10^{-19}~\mathrm{eV} \lesssim m_\phi \lesssim 10^{-10}~\mathrm{eV}$.
The existence of such light scalar fields is motivated by string theory \cite{Arvanitaki:2009fg}.
Ultralight scalar field dark matter starts oscillating when $H \simeq m_\phi$, where $m_\phi$ is its mass \cite{Turner:1983he}.
Since the occupation number is large for our mass range, it can be treated as a classical scalar field in the current universe.

An oscillating classical scalar field causes various peculiar phenomena from which we can probe its existence. 
For instance, the energy-momemtum tensor of the oscillating scalar field generates metric perturbations
that oscillate with half the period of the oscillation of the scalar field. 
\cite{Khmelnitsky:2013lxt} showed that 
such metric perturbations 
can be detected in Pulsar Timing Array observations if the mass of the scalar field is around $10^{-22}~\mathrm{eV}$, 
and null results 
have already put constraints on its abundance \cite{Porayko:2014rfa,Porayko:2018sfa}.
If the scalar field couples with the Standard Model (SM) particles, it mediates additional force \cite{Damour:2010rp}.
Such force is searched for in fifth-force experiments \cite{Adelberger:2003zx} and Equivalence Principle (EP) tests 
such as Lunar Laser Ranging (LLR) experiments \cite{Williams:2004qba,Williams:2005rv}, experiments from E\"ot-Wash group \cite{Schlamminger:2007ht,Wagner:2012ui} and the MICROSCOPE experiment \cite{Touboul:2017grn}.
The oscillation of the scalar field that couples with the SM particles causes time variations 
of physical constants such as the proton mass and the fine-structure constant. 
They produce the time variations of the atomic transitions
as well as oscillatory force on bodies
from which the constraint on the coupling to the SM particles can be obtained \cite{Arvanitaki:2014faa,Arvanitaki:2015iga,Branca:2016rez,Graham:2015ifn,Hees:2016gop,Blas:2016ddr,Arvanitaki:2016fyj,Berge:2017ovy, Hees:2018fpg,Geraci:2018fax}.
For example, hyperfine frequency comparison of ${}^{87}\mathrm{Rb}$ and ${}^{133}\mathrm{Cs}$ atoms has 
been used for this purpose \cite{Hees:2016gop}.

In this paper, under the assumption that an ultralight scalar field comprises all dark matter and it 
weakly couples to the SM particles, we investigate the possibility of detecting such a field with gravitational wave (GW) detectors.
As far as we know, the idea of detecting such a field by the GW detectors was firstly argued by \cite{Arvanitaki:2014faa}. 
Notice that the same idea was discussed in \cite{Pierce:2018xmy} to detect a light vector field.
The main idea is the following.
An ultralight scalar field as dark matter varies spatially on the scale of the de Broglie wavelength 
$k^{-1} \sim 1/(m_\phi \vDM)$ in the Galaxy, 
where $\vDM \sim 10^{-3}$ is the velocity dispersion of dark matter in the Galaxy.
Since the values of the physical constants depend on the value of the scalar field, 
the masses of optical equipments such as mirrors also depend on them.
Then, the spatial variations of the scalar field exerts position-dependent oscillatory force on every optical equipment.
As a result, the optical equipments undergo position-dependent oscillatory motions, which ends up with
non-vanishing signals in the GW detectors' outputs.
Since the frequency of the motion is $f_\phi \simeq m_\phi / 2 \pi$ and the frequency dispersion is tiny, 
$\Delta f_\phi \sim f_\phi \vDM^2$, detectors' signals are nearly monochromatic. 
It is worth mentioning that using the GW detectors to probe the ultralight scalar field dark matter was
also investigated in \cite{Aoki:2016kwl}.
In this reference, not the effects caused by the scalar force but the effects of metric perturbations sourced 
by the scalar field were considered,
and it was found that such effects are far from the detectable level even with the future experiments.
\footnote{Actually, there are a few of flaws in the previous estimate and the constraints can be much tighter. We re-estimate the constraints in Appendix \ref{appendixGR}.}.

To extract information on ultralight scalar field dark matter from real data as much as possible, 
we need to understand its signal's characteristics and develop a suitable data-analysis method.
In this paper, we formulate a data-analysis method toward a detection of ultralight scalar field dark matter with GW detectors.
First, we derive the expression of the signal in an output from a GW detector.
Especially, we take into account detectors' motion due to the Earth's rotation/the detectors' orbital motion around the Sun, which is not taken into account in the previous studies \cite{Arvanitaki:2014faa,Pierce:2018xmy}.
Next, we develop a data-analysis method suitable for the signal.
Taking into account the possibility that we have only one detector, we develop a method which is applicable even with one detector.
Given the sophisticated analysis method, the constraints can be much tighter than the previous estimate.
Finally, we update the previous estimate of the constraints.

The organization of the paper is as follows.
In Section \ref{basics}, we explain the model we consider and how ultralight scalar field dark matter behaves in our Galaxy.
In Section \ref{response}, we obtain a formula for the signal in an output from a gravitational-wave detector.
In Section \ref{detection}, we discuss the signal's characteristics and formulate a suitable data-analysis method to detect this signal.
Then, we update the previous estimate of the constraints with our analysis method.
Section \ref{Conclusion} is devoted to the conclusion.

\section{Basics} \label{basics}
We briefly review the model we consider and how the scalar field dark matter behaves around the gravitational-wave detectors.

\subsection{Model}\label{model}
We consider a scalar field, $\phi$, which couples with particles in the Standard Model and  accounts for all the amount of dark matter. 
We assume that $\phi$ linearly couples with the particles.
Then the low-energy ($\sim 1~\mathrm{GeV}$) effective lagrangian is given by \cite{Damour:2010rp}
\begin{equation}
\mathcal{L}=\mathcal{L}_{\phi}+\mathcal{L}_{\mathrm{SM}} +\mathcal{L}_{\phi-\mathrm{SM}},
\end{equation}
where $\mathcal{L}_{\mathrm{SM}}$ is the Lagrangian of the Standard Model.
$\mathcal{L}_{\phi}$ and $\mathcal{L}_{\phi-\mathrm{SM}}$ are given by
\begin{align}
\mathcal{L}_{\phi}&=-\frac{1}{2} \partial_{\mu} \phi \partial^{\mu} \phi - V(\phi), \\
\mathcal{L}_{\phi-\mathrm{SM}}&=\kappa \phi \left[ \frac{d_e}{4 e^2} F_{\mu \nu} F^{\mu \nu} - \frac{d_g \beta_3}{2 g_3} G^A_{\mu \nu} G^{A \mu \nu} -  \sum_{i=e,u,d} (d_{m_i} + \gamma_{m_i} d_g) m_i \bar{\psi}_i \psi_i \right], \label{L-int}
\end{align}
where $\kappa\equiv \sqrt{4 \pi}/{\Mpl}$, $\beta_3$ is the QCD beta functions and $\gamma_{m_i}$ are the anomalous dimensions of the electrons, the u and d quarks. 
With respect to the coupling between $\phi$ and quarks, it is more convenient to use 
\begin{equation}
d_{\hat{m}}\equiv \frac{d_{m_d} m_d + d_{m_u} m_u}{m_d + m_u},~~~~~~d_{\delta m} \equiv \frac{d_{m_d} m_d - d_{m_u} m_u}{m_d - m_u}
\end{equation}
instead of $d_{m_d}$ and $d_{m_u}$ in the context of equivalence principle tests \cite{Damour:2010rp}.
Since the amplitude of $\phi$ attenuates as the universe expands, we assume that the amplitude is so tiny in the current universe and we can neglect higher-order terms in $V(\phi)$ and approximate it as
\begin{equation}
V(\phi) = \frac{1}{2} m^2_\phi \phi^2.
\end{equation}

\subsection{The scalar field around a detector}  \label{subsec:arounddetector}
We let $(t,~\vec{x})$ be the rest frame of the Solar system and discuss how $\phi$ behaves in this frame.
Dark matter has velocity dispersion of $\vDM \sim 10^{-3}$ in our Galaxy and the Solar system is also moving with the velocity around $\vDM$.
Therefore, the wavenumber of $\phi$, $\vec{k}$, satisfies
\begin{equation}
|\vec{k}| \lesssim m_\phi \vDM.
\end{equation}
Then $\phi$ can be written as
\begin{equation}
\phi = \int d^3 k \left[\phi_{\vec{k}} \e^{i(\omega_{k} t - \vec{k} \cdot \vec{x})} + \cc \right],
\end{equation}
where $|\phi_{\veck}|$ is negligibly small for $|\veck| \gg m_\phi \vDM$.
The angular frequency $\omega_k \sim m_\phi + k^2/2 m_\phi$ and its dispersion $\Delta \omega_k$ are
\begin{equation}
\omega_k \sim m_\phi,~~~~~~~~~\Delta \omega_k \sim m_\phi \vDM^2.
\end{equation}
In addition, for $\phi$ to account for all the amount of dark matter, the following condition must be satisfied,
\begin{equation}
m^2_\phi \left<\phi^2\right> = \rhoDM \simeq 0.3~\mathrm{GeV/cm^3}, \label{localdensity}
\end{equation}
where $<\dots>$ is an average over space with the volume of $V \gg (m_\phi \vDM)^{-3}$ and time of $T \gg m^{-1}_\phi$ and we used a result from \cite{Bovy:2012tw} as a value of $\rhoDM$.

\section{Response of gravitational-wave detectors to scalar waves} \label{response}
As we explained in the previous section, there is a scalar-waves background in the Galaxy if an ultralight scalar field accounts for dark matter.
In this section, we obtain the expression for the signal, $h(t)$, caused by the scalar waves in a GW detector's output.
There are multiple sources of the signal in the output of the detector.
One possible source is a time variation of the laser frequency due to the oscillation of the scalar field.
This mimics common motion of the arms of the detector.
Since interferometric GW detectors are mainly sensitive for their differential motion, 
this effect is subdominant.
The second possible source is modification of laser light's propagation, which is also negligible as we prove in Appendix \ref{propagation}.
Then, we conclude that the signal is mainly sourced by the motion of optical equipments caused by the scalar waves.
In what follows, we first calculate the response of the detectors to monochromatic waves,
\begin{equation}
\phi = \phi_{\veck} \cos (\omega_k t - \vec{k} \cdot \vec{x} + \theta_{\veck}). \label{monochromatic}
\end{equation}
Then we generalize the expression to that for the superposition of the waves.

In our work, we consider Laser Interferometer Gravitational-Wave Observatory (LIGO) \cite{TheLIGOScientific:2014jea}, Einstein Telescope (ET) \cite{Hild:2010id} and Cosmic Explorer (CE) \cite{Evans:2016mbw} as representative ground-based detectors and DECi-hertz Interferometer Gravitational wave Observatory (DECIGO) \cite{Kawamura:2006up} and Laser Interferometer Space Antenna (LISA) \cite{Audley:2017drz} as representative space-based gravitational-wave detectors.
The sensitivity curves for ground-based detectors are taken from \cite{Evans:2016mbw}.
For the DECIGO's sensitivity curve, we use an analytical function in \cite{Yagi:2011wg}.
For LISA, we apply configuration parameters' values summarized in \cite{Audley:2017drz} to calculate the sensitivity.

\subsection{Motion of optical equipments} 
Due to the interaction with $\phi$ given by Eq.~(\ref{L-int}), 
atomic mass depends on the value of $\phi$.
Therefore, the action of an optical equipment in this theory is given by,
\begin{equation}
S=-\int m(\phi) \sqrt{-\eta _{\mu \nu} dx^\mu dx^\mu},
\end{equation}
where we neglect gravitational fields.
Then the equation of motion of the optical equipment in the non-relativistic limit becomes
\begin{equation}
\frac{d^2 x^i}{d t^2} \approx - \kappa \alpha(\phi) \left( \partial_i \phi+
{\dot \phi} \frac{dx^i}{dt} \right),
~~~~~~~~~\alpha(\phi) \equiv \frac{d \mathrm{ln} m(\phi)}{d (\kappa \phi)}.
\end{equation}
Assuming that the equipment is at rest in the absence of $\phi$, 
the second term on the right hand side is second order in the amplitude
of the scalar field and is subdominant compared to the first term.
Thus, the equation of motion becomes
\begin{equation}
\frac{d^2 x^i}{d t^2} \simeq - \kappa \alpha(\phi) \partial_i \phi. \label{EOM}
\end{equation}
Since the masses of atoms are mostly determined by the QCD energy scale, $\alpha(\phi)$ is approximately given by \cite{Damour:2010rp}
\begin{equation}
\alpha(\phi) \simeq d^*_g \simeq d_g + 0.093 (d_{\hat{m}} - d_g).  \label{alpha}
\end{equation}
Substituting Eq. (\ref{monochromatic}) and Eq. (\ref{alpha}) into Eq. (\ref{EOM}), we obtain
\begin{equation}
\frac{d^2 x^i}{d t^2} \simeq -d^*_g \kappa \phi_{\veck} k^i \sin(\omega_k t - \veck \cdot \vec{x}_0 + \theta_{\veck}),
\end{equation}
where $\vec{x}_0$ is the position on which the equipment is in the absence of $\phi$.
Solving this leads to
\begin{equation}
x^i \simeq d^*_g \kappa \phi_{\veck} \frac{k^i}{m^2_\phi} \sin(\omega_k t - \veck \cdot \vec{x}_0 + \theta_{\veck}) + \mathrm{const}.
\end{equation}

\subsection{Response of the interferometric gravitational-wave detectors}
Let us next derive the expression of the signal caused by the motion of the optical equipments.
First, we calculate the response of a Michelson-interferometer with arms along the unit vectors, $\vec{n}$ and $\vec{m}$.
Following the method used in \cite{Rakhmanov:2008is}, we calculate the time necessary for photon to make a round trip through the arms.
The oscillation of the mirror at $\vec{x}$ due to the scalar field is given by
\begin{equation}
\delta x^i(t,\vec{x}) \equiv d^*_g \kappa \phik \frac{k^i}{m^2_\phi} \sin(\omega_k t - \veck \cdot \vec{x} + \thetak).
\end{equation}
Then the position of the front mirror, $\vec{x}_1$, and that of the end mirror, $\vec{x}_2$, in the arm along $\vec{n}$ are
\begin{align}
x^i_1(t)&=x^i + \delta x^i(t,\vec{x}),\\
x^i_2(t)&=x^i+L n^i + \delta x^i(t,\vec{x} + L \vec{n}),
\end{align}
where $L$ is the arm length in the absence of the scalar field.
Therefore the perturbation of the round-trip time due to the oscillation, $\delta t(t; L, \vec{n})$, is
\begin{equation}
\delta t(t;L,\vec{n}) \simeq n_i ( -\delta x^i(t,\vec{x}) + 2 \delta x^i (t-L,\vec{x}+L \vec{n}) - \delta x^i (t- 2 L,\vec{x})),
\end{equation}
where $t$ is the time where photon returns back to the front mirror and higher order terms  in $\delta x$ have been neglected.
This can be separated into two parts,
\begin{equation}
\delta t(t;L,\vec{n}) = n_i ( -\delta x^i(t,\vec{x}) + 2 \delta x^i (t-L,\vec{x}) - \delta x^i (t- 2 L,\vec{x})) + 2 L n_i n^j \partial_j \delta x^i (t-L,\vec{x}). \label{derivative-exp}
\end{equation}
The first part is
\begin{align}
\delta t_1 (t;L,\vec{n}) &\equiv n_i ( -\delta x^i(t,\vec{x}) + 2 \delta x^i (t-L,\vec{x}) - \delta x^i (t- 2 L,\vec{x})) \nonumber \\
&\simeq 4 d^*_g \kappa \phik \frac{\vec{k} \cdot \vec{n}}{m^2_\phi} \sin^2\left(\frac{m_\phi L}{2}\right) \sin(\omega_k (t-L) - \vec{k}\cdot\vec{x}+\thetak),
\end{align}
and the second part is
\begin{align}
\delta t_2 (t;L,\vec{n}) &\equiv 2 L n_i n^j \partial_j \delta x^i (t-L,\vec{x}) \nonumber \\
&\simeq -2 d^*_g \kappa \phik L \frac{(\veck \cdot \vec{n})^2}{m^2_\phi} \cos (\omega_k (t-L) - \veck \cdot \vec{x} + \thetak).
\end{align}
Since $|\vec{k} \cdot \vec{n}| \sim m_\phi \vDM$, the round-trip time can be approximated as follows,
\begin{equation}
\delta t(t;L,\vec{n}) \simeq \begin{cases}
\delta t_1 (t;L,\vec{n}) & \left( 2 \frac{\sin^2\left(\frac{m_\phi L}{2} \right)}{m_\phi L} > \vDM \right) \\
\delta t_2 (t;L,\vec{n})  & \left( 2 \frac{\sin^2\left(\frac{m_\phi L}{2} \right)}{m_\phi L} < \vDM \right)
\end{cases}
\end{equation}
From this, we can calculate the phase delay of the laser light after a round-trip as follows,
\begin{equation}
\varphi (t;L,\vec{n}) = 2 \pi f_{\mathrm{laser}}(2 L + \delta t(t;L,\vec{n})),
\end{equation}
where $f_{\mathrm{laser}}$ is the frequency of the laser light. 
The derivative expansion in Eq.~(\ref{derivative-exp}) breaks down
for $L > 1/(m_\phi \vDM)$. This situation never realizes for any detectors
considered in this paper.
Phase delay for another arm is obtained by replacing $\vec{n}$ by $\vec{m}$.
Since the signal in the gravitational-wave channel is the difference of the phases of the laser light traveling in the two arms, it is
\begin{equation}
h(t) = \frac{\varphi(t; L,\vec{m}) - \varphi(t;L,\vec{n})}{4 \pi f_{\mathrm{laser}} L} \simeq \begin{cases}
h_1 (t) & \left( 2 \frac{\sin^2\left(\frac{m_\phi L}{2} \right)}{m_\phi L} > \vDM \right) \\
h_2 (t)  & \left( 2 \frac{\sin^2\left(\frac{m_\phi L}{2} \right)}{m_\phi L} < \vDM \right)
\end{cases}
\end{equation}
where
\begin{align}
h_1(t)&=2 d^*_g \kappa \phik \frac{\sin^2\left(\frac{m_\phi L}{2}\right)}{m_\phi L} \frac{\veck \cdot \vec{n}-\veck \cdot \vec{m}}{m_\phi} \sin(\omega_k (t-L)-\veck \cdot \vec{x} + \thetak), \\
h_2(t)&=-d^*_g \kappa \phik \frac{(\veck \cdot \vec{n})^2 - (\vec{k} \cdot \vec{m})^2}{m^2_\phi} \cos (\omega_k (t-L) - \veck \cdot \vec{x} + \thetak).
\end{align}
Generalizing this expression into that for the superposition of the non-relativistic waves, we obtain
\begin{align}
h_1(t)&=2 d^*_g \kappa  \frac{\sin^2\left(\frac{m_\phi L}{2}\right)}{m^2_\phi L} (n^i - m^i)\partial_i \phi(t,\vec{x}), \\
h_2(t)&=d^*_g \kappa \frac{n^i n^j - m^i m^j}{m^2_\phi} \partial_i \partial_j \phi(t,\vec{x}) .
\end{align}
Taking into account the time variations of the detectors' orientation, we must replace $\vec{n}$ and $\vec{m}$ by time-varying vectors, $\vec{n}(t)$ and $\vec{m}(t)$.
For ground-based detectors, the time variation is due to the Earth's rotation.
Then the expression for the signal is
\begin{align}
h_1(t)&=2 d^*_g \kappa  \frac{\sin^2\left(\frac{m_\phi L}{2}\right)}{m^2_\phi L} \left( g_{1,0}(t)+g_{1,1}(t) \e^{i \Od t} + g_{1,1}^*(t) \e^{-i \Od t} \right), \label{hhg} \\ 
h_2(t)&=d^*_g \frac{\kappa}{m^2_\phi} \left(g_{2,0}(t) + g_{2,1}(t) \e^{i \Od t} + g_{2,1}^*(t) \e^{-i \Od t} + g_{2,2}(t) \e^{2 i \Od t} + g_{2,2}^*(t) \e^{-2 i \Od t}\right), \label{hlg}
\end{align}
where $\Od=2 \pi/\text{(sidereal day)}$.
For the space-based detector, DECIGO, the expression is
\begin{align}
h_1(t)=&2 d^*_g \kappa  \frac{\sin^2\left(\frac{m_\phi L}{2}\right)}{m^2_\phi L}  \left(c_{1,0}(t) + c_{1,1}(t) \e^{i \Oy t} + c_{1,1}^*(t) \e^{-i \Oy t} + c_{1,2}(t) \e^{2 i \Oy t} + c_{1,2}^*(t) \e^{-2 i \Oy t}\right), \label{hhd}\\
h_2(t)=&d^*_g \frac{\kappa}{m^2_\phi}  (c_{2,0}(t) + c_{2,1}(t) \e^{i \Oy t} + c_{2,1}^*(t) \e^{-i \Oy t} + c_{2,2}(t) \e^{2 i \Oy t} + c_{2,2}^*(t) \e^{-2 i \Oy t} \nonumber \\
&~~~~~~~~~~~~~+c_{2,3}(t) \e^{3 i \Oy t} + c_{2,3}^*(t) \e^{-3 i \Oy t} + c_{2,4}(t) \e^{4 i \Oy t} + c_{2,4}^*(t) \e^{-4 i \Oy t} ) \label{hld},
\end{align}
where $\Oy=2 \pi/\text{year}$.
The functions, $g_{\mathrm{1},i}(t)$, $g_{\mathrm{2},i}(t)$, $c_{\mathrm{1},i}(t)$ and $c_{\mathrm{2},i}(t)$, have a frequency range of  $f_\phi < f \lesssim f_\phi (1+\vDM^2)$, where $f_\phi \equiv m_\phi/2 \pi$.
Their expressions are summarized in Appendix. \ref{timevariation}.

Next we calculate the response of LISA. 
There are several methods to cancel out laser-frequency noise in the LISA experiment and a measured variable depends on which method we choose.
In this paper, we apply the second-generation TDI variable $X_1(t)$ \cite{Shaddock:2003dj, Cornish:2003tz}.
Note that the Doppler shift caused by scalar waves can be calculated by
\begin{equation}
y^{\mathrm{SW}}(t; L, \vec{n}) \equiv -\frac{1}{2 \pi f_{\mathrm{laser}}} \frac{d \varphi(t;L,\vec{n})}{d t} 
\end{equation}
and the contribution from scalar waves to $X_1(t)$ is approximately
\begin{align}
X^{\mathrm{SW}}_1(t) \simeq& y(t;L,\vec{n})- y(t;L,\vec{m})-y(t-2 L;L,\vec{n}) +y(t-2L;L.\vec{m}) \nonumber \\
& -y(t-4 L;L,\vec{n})+ y(t-4L;L,\vec{m})+y(t-6 L;L,\vec{n}) -y(t-6L;L.\vec{m}) ,
\end{align}
where $\vec{n}$ and $\vec{m}$ are unit vectors along two different arms of LISA and $L$ is the arm length of LISA in this case.
Then it is straightforward to calculate $X^{\mathrm{SW}}_1(t)$ and the signal is
\begin{equation}
h(t)=X^{\mathrm{SW}}_1(t) = \begin{cases}
X_{1,1} (t) & \left( 2 \frac{\sin^2\left(\frac{m_\phi L}{2} \right)}{m_\phi L} > \vDM \right) \\
X_{1,2} (t)  & \left( 2 \frac{\sin^2\left(\frac{m_\phi L}{2} \right)}{m_\phi L} < \vDM \right),
\end{cases}
\end{equation}
where 
\begin{align}
X_{1,1}(t) &\simeq -\frac{16 d^*_g \kappa}{m^2_\phi} \sin^2\left(\frac{m_\phi L}{2}\right) \sin (m_\phi L) \sin(2 m_\phi L) (n^i - m^j) \partial_0 \partial_i \phi(t,\vec{x}), \\
X_{1,2}(t) & \simeq -8 d^*_g \kappa L \sin(m_\phi L) \sin(2 m_\phi L) \frac{n^i n^j - m^i m^j}{m^2_\phi} \partial_0 \partial_i \partial_j \phi(t,\vec{x}). 
\end{align}
Taking into account the time variations of the detector's orientation, 
the expression of the signal becomes
\begin{align}
X_{1,1}(t)=&-16 d^*_g \kappa \sin^2\left(\frac{m_\phi L}{2}\right) \sin (m_\phi L) \sin(2 m_\phi L) \nonumber \\
&~~~~~~~~~~~~\times  \left(x_{1,0}(t) + x_{1,1}(t) \e^{i \Oy t} + x_{1,1}^*(t) \e^{-i \Oy t} + x_{1,2}(t) \e^{2 i \Oy t} + x_{1,2}^*(t) \e^{-2 i \Oy t}\right),  \label{Xh}  \\
X_{1,2}(t)=&-8 d^*_g \kappa L \sin(m_\phi L) \sin(2 m_\phi L)  (x_{2,0}(t) + x_{2,1}(t) \e^{i \Oy t}+ x_{2,1}^*(t) \e^{-i \Oy t} + x_{2,2}(t) \e^{2 i \Oy t} \nonumber \\
&~~~~ + x_{2,2}^*(t) \e^{-2 i \Oy t} +x_{2,3}(t) \e^{3 i \Oy t} + x_{2,3}^*(t) \e^{-3 i \Oy t} + x_{2,4}(t) \e^{4 i \Oy t} + x_{2,4}^*(t) \e^{-4 i \Oy t} ). \label{Xl}
\end{align}
The functions, $x_{\mathrm{1},i}(t)$ and $x_{\mathrm{2},i}(t)$, have a frequency range of  $f_\phi < f \lesssim f_\phi (1+\vDM^2)$.
Their expressions are also summarized in Appendix \ref{timevariation}.

\section{Detection method and future constraints} \label{detection}
In this section, we discuss an appropriate detection method for the signal
derived in the previous section and the future constraints obtained by this method.
In this section, we use $f_\phi = m_\phi/2 \pi$ instead of $m_\phi$.

\subsection{Characterization of the signal} \label{characterization}
\begin{figure}
  \begin{center}
    \begin{tabular}{c}

      \begin{minipage}{0.5\hsize}
        \begin{center}
          \includegraphics[clip, width=8cm]{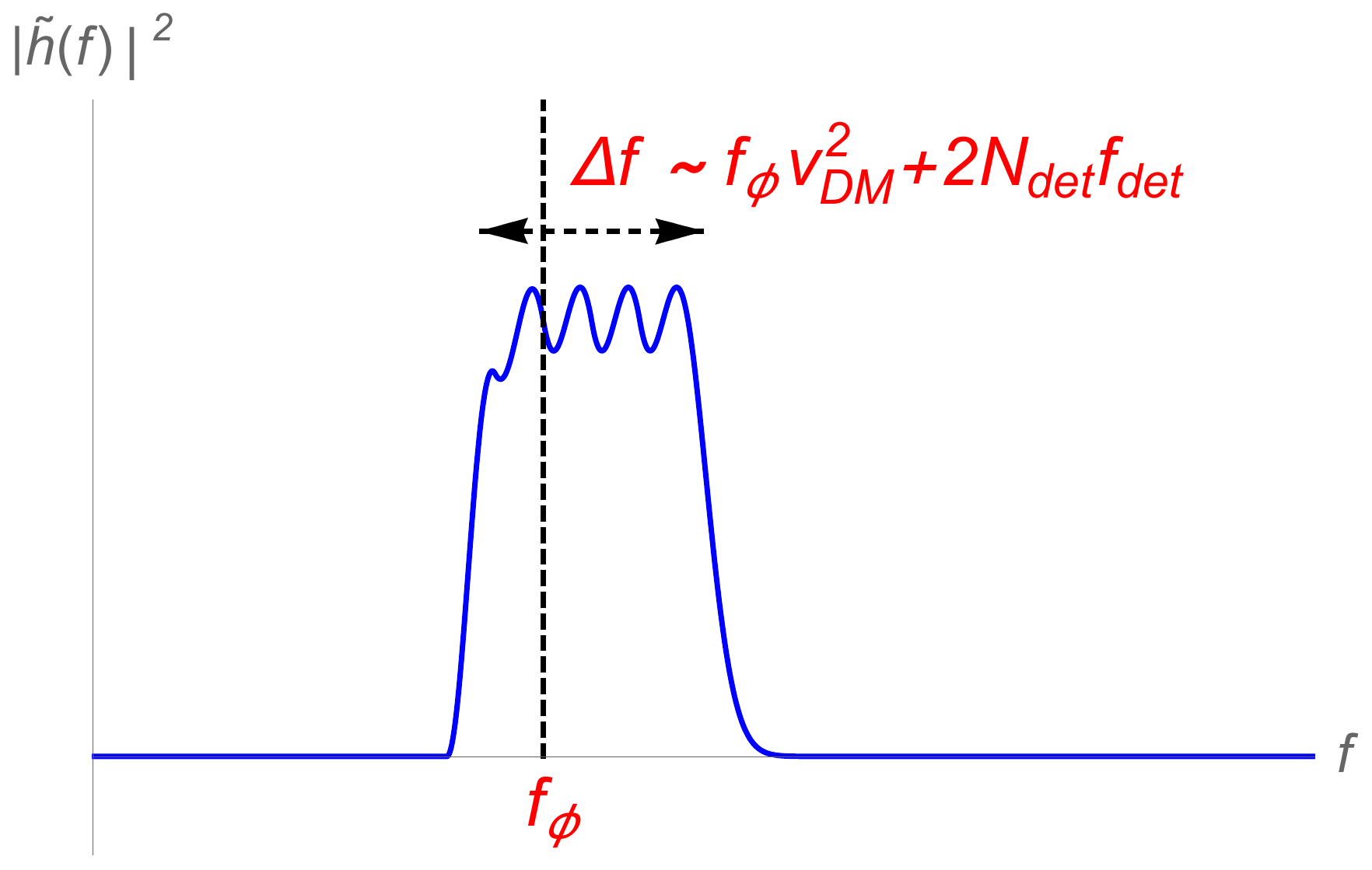}
        \end{center}
      \end{minipage} 

      \begin{minipage}{0.5\hsize}
 	      \begin{center}
    	     \includegraphics[clip, width=8cm]{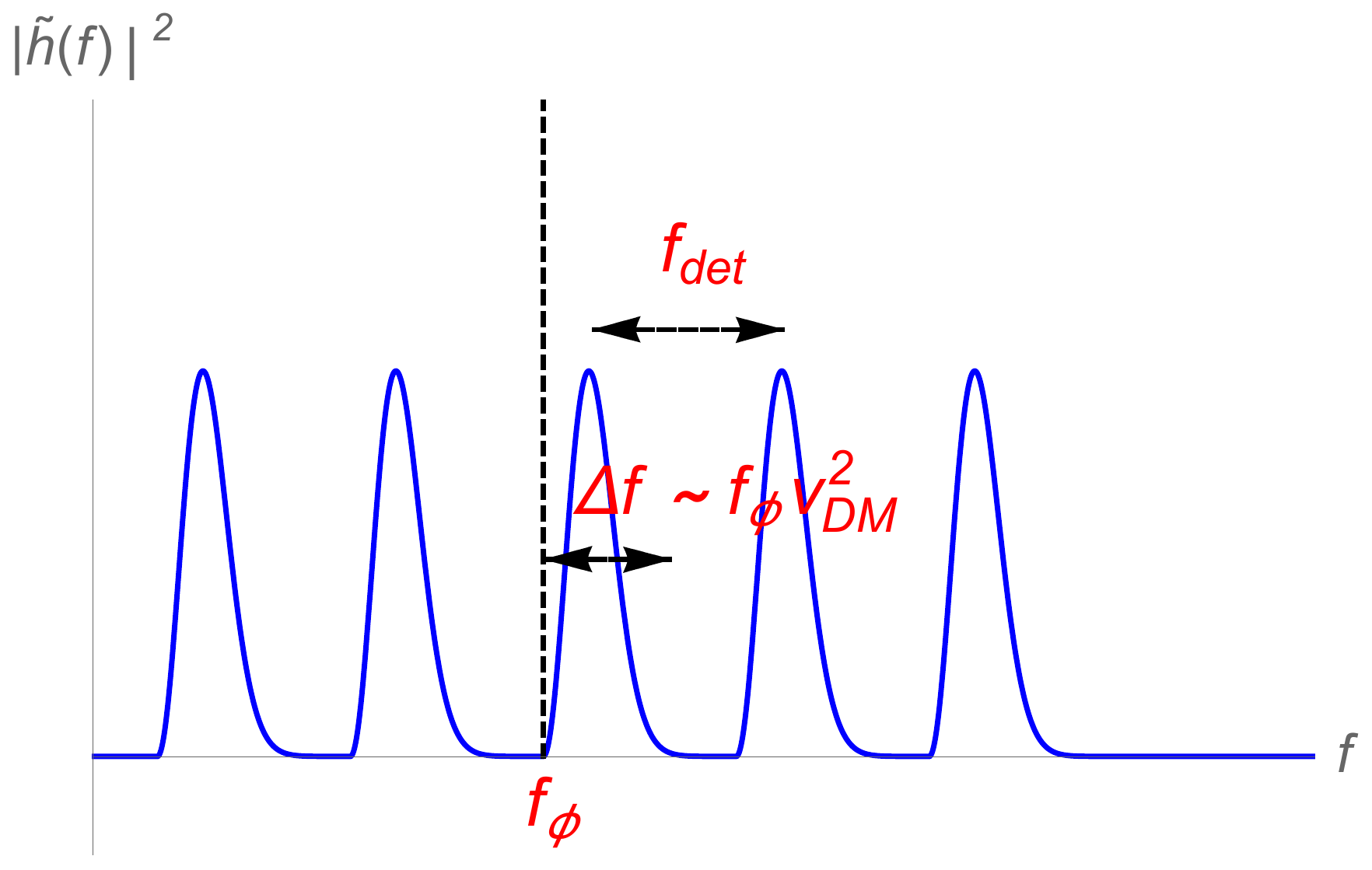}
        	\end{center}
      	\end{minipage}
      
    \end{tabular}
    \caption{The schematic picture of the signal's spectra. The left one is for $f_\phi \vDM^2 \gtrsim \fdet$ and the right one for $f_\phi \vDM^2 \ll \fdet$. } 
    \label{spectrum}
  \end{center}
\end{figure}

Before moving to the detection method, we briefly summarize the characterization of the signal.
Looking at Eq. (\ref{hhg}), (\ref{hlg}), (\ref{hhd}), (\ref{hld}), (\ref{Xh}) and (\ref{Xl}), we find that the signal is a sum of $2 \Ndet (f_\phi)+1$ spectra with width of $\sim f_\phi \vDM^2$, where 
\begin{equation}
\Ndet (f_\phi)=
 \begin{cases}
1  & \left( \frac{\sin^2\left(\pi f_\phi L \right)}{\pi f_\phi L} > \vDM \right) \\
2  & \left( \frac{\sin^2\left(\pi f_\phi L \right)}{\pi f_\phi L} < \vDM \right)
\end{cases}
\end{equation}
for ground-based detectors and
\begin{equation}
\Ndet (f_\phi)=
 \begin{cases}
2  & \left( \frac{\sin^2\left(\pi f_\phi L \right)}{\pi f_\phi L} > \vDM \right) \\
4  & \left( \frac{\sin^2\left(\pi f_\phi L \right)}{\pi f_\phi L} < \vDM\right)
\end{cases}
\end{equation}
for space-based detectors.
Each spectrum has frequency band of $f_\phi + j \fdet < f < f_\phi (1 + \vDM^2) + j \fdet$ ($j=-\Ndet,-\Ndet+1,\dots,\Ndet$), where
\begin{equation}
\fdet \equiv
 \begin{cases}
\Od/2 \pi  & \left( \text{ground-based detectors} \right) \\
\Oy/2 \pi  & \left( \text{space-based detectors} \right).
\end{cases}
\end{equation}
Therefore, if $f_\phi \vDM^2 \gtrsim \fdet$ the signal has one broadened spectrum whose frequency band is $f_\phi - \Ndet (f_\phi) \fdet < f < f_\phi (1 + \vDM^2) + \Ndet (f_\phi) \fdet$, and if $f_\phi \vDM^2 < \fdet$ the signal has $2 \Ndet (f_\phi)+ 1$ separated spectra.
The schematic spectrum is drawn in Fig. \ref{spectrum}.

\subsection{Detection method}

Next we discuss an appropriate method to detect this signal.
The data from a GW detector is a sum of the signal, $h(t)$, and the noise, $n(t)$,
\begin{equation}
s(t) = h(t) + n(t).
\end{equation}
Here we assume that the noise is stationary Gaussian.
The Fourier components of the data are
\begin{equation}
\tilde{s}(f_k) = \int^{\Tobs}_0 s(t) \e^{-2 \pi i f_k t} dt,~~~~~f_k = \frac{k}{\Tobs},
\end{equation}
where $\Tobs$ is the observation time.
From the discussions in \ref{characterization}, the signal's power is in the following frequency bins,
\begin{align}
&F(f_\phi) = \bigcup^{\Ndet (f_\phi)}_{ j = - \Ndet (f_\phi)} F_j(f_\phi), \\
&F_j(f_\phi) = \left\{ f_k ~|~ R\left((f_\phi + j \fdet) \Tobs \right) \leq k \leq R\left( \left( f_\phi (1 + \vDM^2) + j \fdet \right) \Tobs \right),~k \in  \mathbb{N} \right\},
\end{align}
where $R(x)$ is the integer closest to $x$.
In what follows, we present two efficient ways to extract this signal from the noise.

\subsubsection{Single detector: Incoherent sum of spectra over $F(f_\phi)$}


One simple way is to take a sum of spectra over the signal's frequency range.
The detection statistic in this method is
\begin{equation}
\rho(f_\phi) \equiv \sum_{f_k \in F(f_\phi)} \frac{2 |\tilde{s}(f_k)|^2}{\Tobs S(f_k)},
\end{equation}
where $S(f)$ is the one-sided power spectrum of the noise.
Each term in the sum is normalized so that its expectation value is 1 if the data is pure noise.


The threshold of $\rho(f_\phi)$ to claim detection is determined by its statistical properties.
If the data is pure noise, the cumulative distribution function for $\rho(f_\phi)$ is
\begin{equation}
\mathrm{CDF}\left[\rho(f_\phi) < \rhoc \right] = \frac{\gamma(N(f_\phi), \rhoc)}{(N(f_\phi)-1)!},
\end{equation}
where $N(f_\phi)$ denotes the number of elements in $F(f_\phi)$.
Given a false alarm probability, $F$, the threshold $\rhoc$ is determined by
\begin{equation}
1 - \mathrm{CDF}\left[\rho(f_\phi) < \rhoc\right] = F.
\end{equation}
Since the search is performed for various values of $f_\phi$, the number of the trials, $\Nt$, should be taken into account when $F$ is determined.
Assuming that the search is performed over the detector's frequency range at intervals of $\sim 1/\Tobs$, the trial number is $(f_{\mathrm{max}} - f_{\mathrm{min}}) \Tobs + 1$, where $f_{\mathrm{min}}$ and $f_{\mathrm{max}}$ are minimum and maximum frequency of the frequency range over which the search is taken over.
For example, to reject the existence of the signal with 2 sigma level, $F$ should be
\begin{equation}
F \simeq \frac{0.05}{\Nt}. \label{FAP}
\end{equation}
The threshold of the signal's amplitude which satisfies $E[\rho(f_\phi)] = \rhoc$ is
\begin{equation}
\hth(f_\phi) =  \sqrt{ \frac{S(f_\phi)}{2 \Tobs} \left( \rhoc - N(f_\phi) \right) }.
\end{equation}

\subsubsection{Two detectors: Narrow band stochastic gravitational-wave background search}


Another way is to take correlation between the outputs from two detectors, 
which is used to search for stochastic GW background \cite{Allen:1997ad,TheLIGOScientific:2016dpb} and considered in the previous study on search for a light vector field \cite{Pierce:2018xmy}.
The detection statistic is
\begin{equation}
\rho(f_\phi) = \frac{1}{\Tobs} \sum^{\infty}_{k=-\infty} \tilde{s}^*_{1}(f_k) \tilde{s}_{2}(f_k) \tilde{Q}(f_k; f_\phi),
\end{equation}
where $\tilde{s}_1(f_k)$ and $\tilde{s}_2(f_k)$ are Fourier components of the data from two detectors.
$\tilde{Q}(f_k; f_\phi)$ should be determined so that the search is optimal.
Since the wavelength of the scalar field, $1/m_\phi \vDM$, is much longer than that of GWs with the same frequency, the overlap reduction function \cite{Allen:1997ad} is approximately constant.
In addition, we only assume the signal's bandwidth and do not make any assumptions on its spectral shape.
Under these considerations, we apply the following filter function,
\begin{equation}
\tilde{Q}(f_k; f_\phi) = \frac{1}{\sqrt{S_1(f_k) S_2(f_k)}} \Theta(f_k; f_\phi),
\end{equation}
where
\begin{equation}
\Theta(f_k; f_\phi) = \begin{cases}
1 & (|f_k| \in F(f_\phi)) \\
0 & (\text{otherwise}) \end{cases}
\end{equation}\\
\par
Having explained the two detection methods, let us
estimate the sensitivities of these methods.
For the former method, the deviation of the expectation value of $\rho(f_\phi)$ due to the presence of the signal is
\begin{equation}
E[\rho(f_\phi)]_{s(t)=h(t)+n(t)} - E[\rho(f_\phi)]_{s(t)=n(t)} \simeq \frac{\Tobs}{S(f_\phi)} \overline{h^2},
\end{equation}
where
\begin{equation}
\overline{h^2} = \frac{1}{\Tobs}\int^{\Tobs}_0 h^2(t) dt.
\end{equation}
On the other hand, its variance when the data is pure noise is
\begin{equation}
E\left[\left(\rho(f_\phi) - E[\rho(f_\phi)]\right)^2\right]  = 2 N(f_\phi).
\end{equation}
Therefore, the threshold of $\sqrt{\overline{h^2}}$ for detection is
\begin{equation}
\hth(f_\phi) \sim \frac{\left( 2 N(f_\phi) \right)^{\frac{1}{4}}}{\sqrt{\Tobs}} \sqrt{S(f_\phi)}. \label{threshold_incoherent}
\end{equation}

Next, we estimate the sensitivity of the latter method assuming that two detectors are parallel and their sensitivity curves are the same.
The expectation value of the detection statistic when the signal is present is
\begin{equation}
E[\rho(f_\phi)] \simeq \frac{\Tobs}{S(f_\phi)} \overline{h^2},
\end{equation}
while its variance when data is pure noise is
\begin{equation}
E\left[\left(\rho(f_\phi) - E[\rho(f_\phi)]\right)^2\right]  = \frac{N(f_\phi)}{2}.
\end{equation}
Therefore, the threshold amplitude for detection is
\begin{equation}
\hth (f_\phi) \sim \frac{N^{\frac{1}{4}}(f_\phi)}{2^{\frac{1}{4}} \sqrt{\Tobs}} \sqrt{S(f_\phi)},
\end{equation}
which is different from Eq. (\ref{threshold_incoherent}) only by a $\mathcal{O}(1)$ factor.
Since the former method requires only one detector, 
we focus on that method and estimate the constraints given by experiments with gravitational-wave detectors in this paper.

Especially in the case where $\Tobs > 1/f_\phi \vDM^2$, $N(f_\phi) \sim f_\phi \vDM^2 \Tobs$ and
\begin{equation}
\hth(f_\phi) \sim \frac{\sqrt{f_\phi \vDM^2}}{N^{\frac{1}{4}} (f_\phi)} \sqrt{S(f_\phi)}. \label{threshold_incoherent2}
\end{equation}
Note that in addition to improvement by a factor of $(1/f_\phi \vDM^2)^{1/2}$ mentioned in \cite{Arvanitaki:2014faa}, there is another improvement by a factor of 
$N^{1/4}(f_\phi)$, which is $\sim 7$ for $f_\phi=100~\mathrm{Hz}$ and $\Tobs = 1~\mathrm{year}$.
The latter factor is overloooked in the literature,
and our estimate improves the previous estimate due to this factor.

\subsection{Future constraints}

\begin{figure}
	\begin{center}
		\includegraphics[width = 12cm]{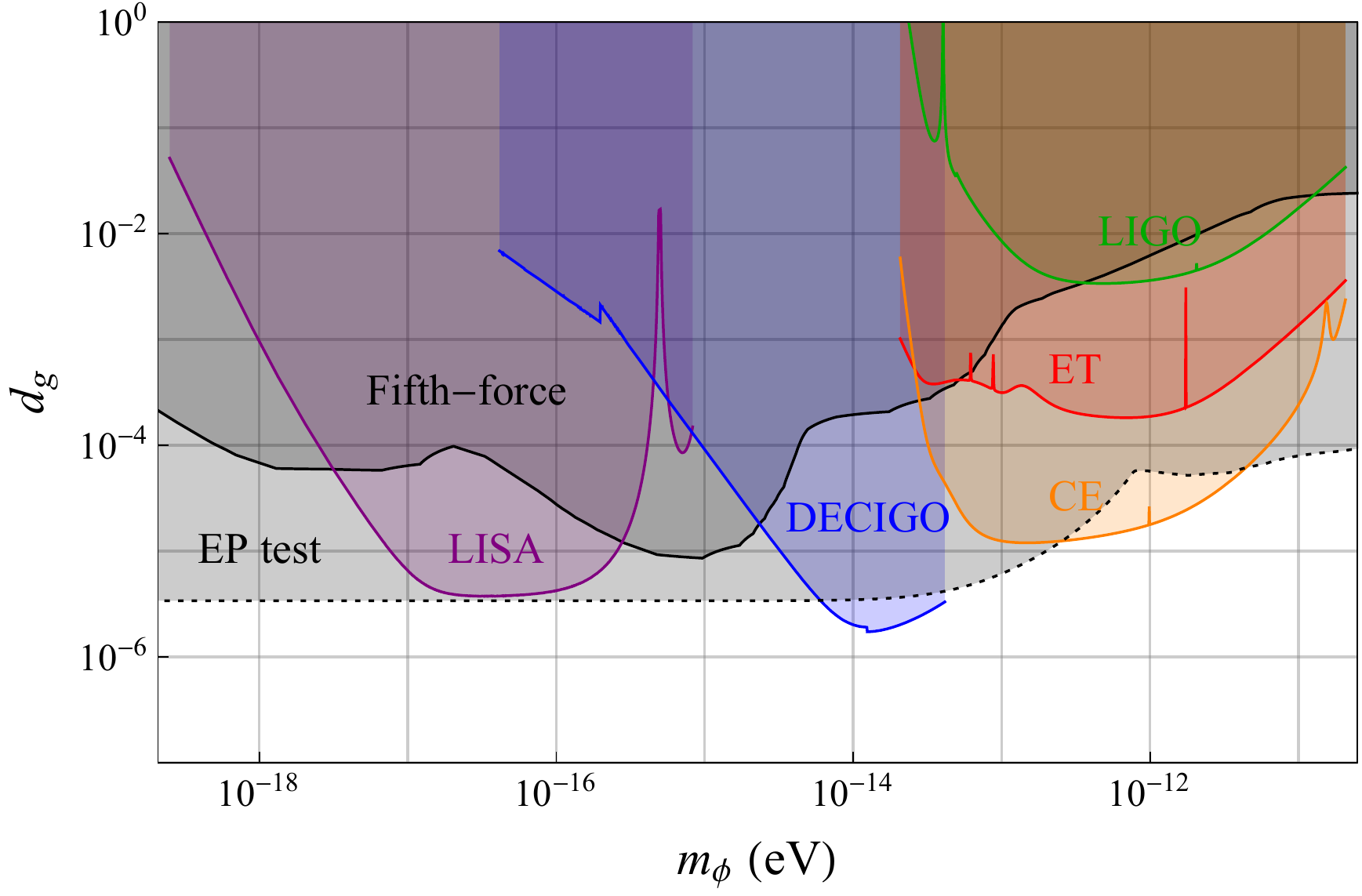}
		\caption{The future constraints we can obtain with gravitational-wave detectors and the constraints given by the fifth-force experiments (Fifth-force) and the tests of equivalence principle (EP test) are shown. The shaded region is or will be excluded. We consider the case where the coupling constants other than $d_g$ are zero here. We assume that $\Tobs = 1~\mathrm{year}$ and all of them are $2 \sigma$ limits. The constraints of EP tests shown here are given by an experiment from E\"ot-Wash group \cite{Wagner:2012ui} and MICROSCOPE \cite{Touboul:2017grn}.  The constraints given by LLR experiments \cite{Williams:2004qba,Williams:2005rv} and a hyperfine frequency comparison experiment \cite{Hees:2016gop} are less stringent in this frequency range. Note that the constraints from EP tests depend on the values of the other coupling constants and they give no constraints if $d_g = d_{m_e} = d_{m_u} = d_{m_d}$ and $d_e = 0$.}
	\label{constraints}
	\end{center}
\end{figure}

Next we estimate the future constraints on this dark matter model with the former detection method described in the previous subsection.
For simplicity, we consider a plane wave satisfying (\ref{localdensity}),
\begin{equation}
\phi=\frac{\sqrt{2 \rhoDM}}{m_\phi} \cos(\omegak t - \veck \cdot \vec{x} + \thetak),~~~~|\veck| = m_\phi \vDM. \label{Monotonic}
\end{equation}
Averaging the square of the signal over the directions of $\veck$ and time, we obtain
\begin{equation}
\overline{h^2}(f_\phi) = \displaystyle \lim_{T \to \infty} \frac{1}{T} \int^{T/2}_{-T/2} dt \int \frac{d \Omega_{\vec{k}}}{4 \pi} h^2(t) = 
\begin{cases}
\frac{8}{3}  \vDM^2 d^2_g \frac{\kappa^2 \rhoDM}{m^4_\phi L^2} (1- \vec{n} \cdot \vec{m}) \sin^4\left(\frac{m_\phi L}{2}\right) & \left( 2 \frac{\sin^2\left(\frac{m_\phi L}{2} \right)}{m_\phi L} > \vDM \right) \\
\frac{4}{15}  \vDM^4 d^2_g \frac{\kappa^2 \rhoDM}{m^2_\phi} \left(1- (\vec{n} \cdot \vec{m})^2\right)  & \left( 2 \frac{\sin^2\left(\frac{m_\phi L}{2} \right)}{m_\phi L} < \vDM \right)
\end{cases}
\end{equation}
for detectors other than LISA and
\begin{equation}
\overline{h^2}(f_\phi) = 
\begin{cases}
\frac{256}{3} \vDM^2 d^2_g \frac{\kappa^2 \rhoDM}{m^2_\phi} \sin^4 \left(\frac{m_\phi L}{2}\right) \sin^2(m_\phi L) \sin^2(2 m_\phi L) & \left( 2 \frac{\sin^2\left(\frac{m_\phi L}{2} \right)}{m_\phi L} > \vDM \right) \\
\frac{64}{5} \vDM^4 d^2_g \kappa^2 L^2 \rhoDM \sin^2 (m_\phi L) \sin^2 (2 m_\phi L) & \left( 2 \frac{\sin^2\left(\frac{m_\phi L}{2} \right)}{m_\phi L} < \vDM \right)
\end{cases}
\end{equation}
for LISA.
Then, the criteria, $\sqrt{\overline{h^2}(f_\phi)} > \hth(f_\phi)$, leads to the parameter region of $d_g$ we will be able to probe with gravitational-wave detectors.
Fig. \ref{constraints} compares this parameter region and constraints given from the fifth-force experiments \cite{Adelberger:2003zx} and tests of equivalence principle (EP tests) \cite{Wagner:2012ui, Touboul:2017grn}.
All the detectors will provide more stringent constraints than the fifth-force experiments do in a mass band.
Especially, the constraints at $m_\phi = 2 \times 10^{-17}~\mathrm{eV},~10^{-14}~\mathrm{eV}$ and $10^{-12}~\mathrm{eV}$ are improved by a factor of $\sim 30$, $\sim 100$ and $\sim 350$ respectively.

\begin{figure}
  \begin{center}
    \begin{tabular}{c}

      \begin{minipage}{0.5\hsize}
        \begin{center}
          \includegraphics[clip, width=8cm]{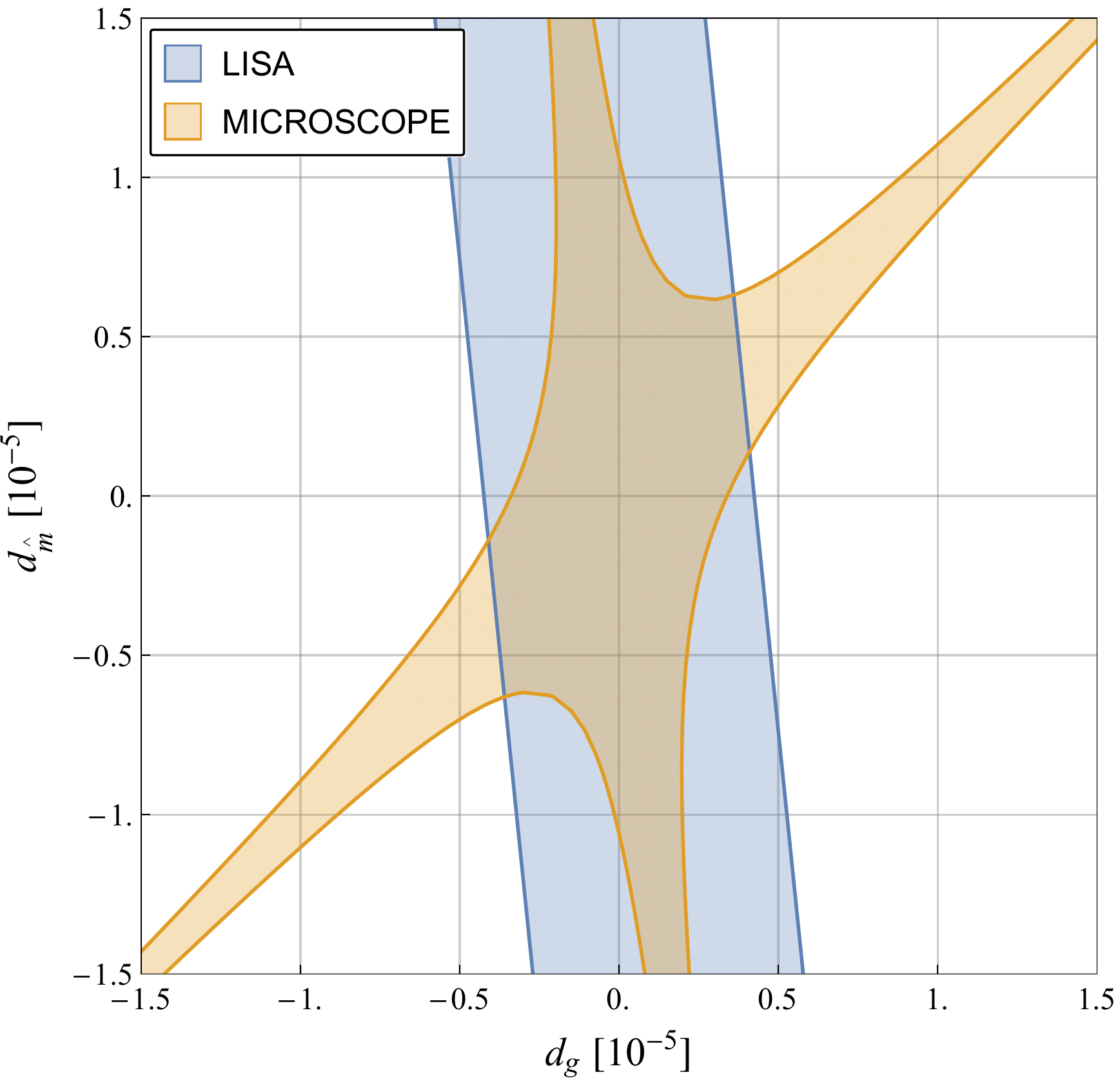}
        \end{center}
      \end{minipage} 

      \begin{minipage}{0.5\hsize}
 	      \begin{center}
    	     \includegraphics[clip, width=8cm]{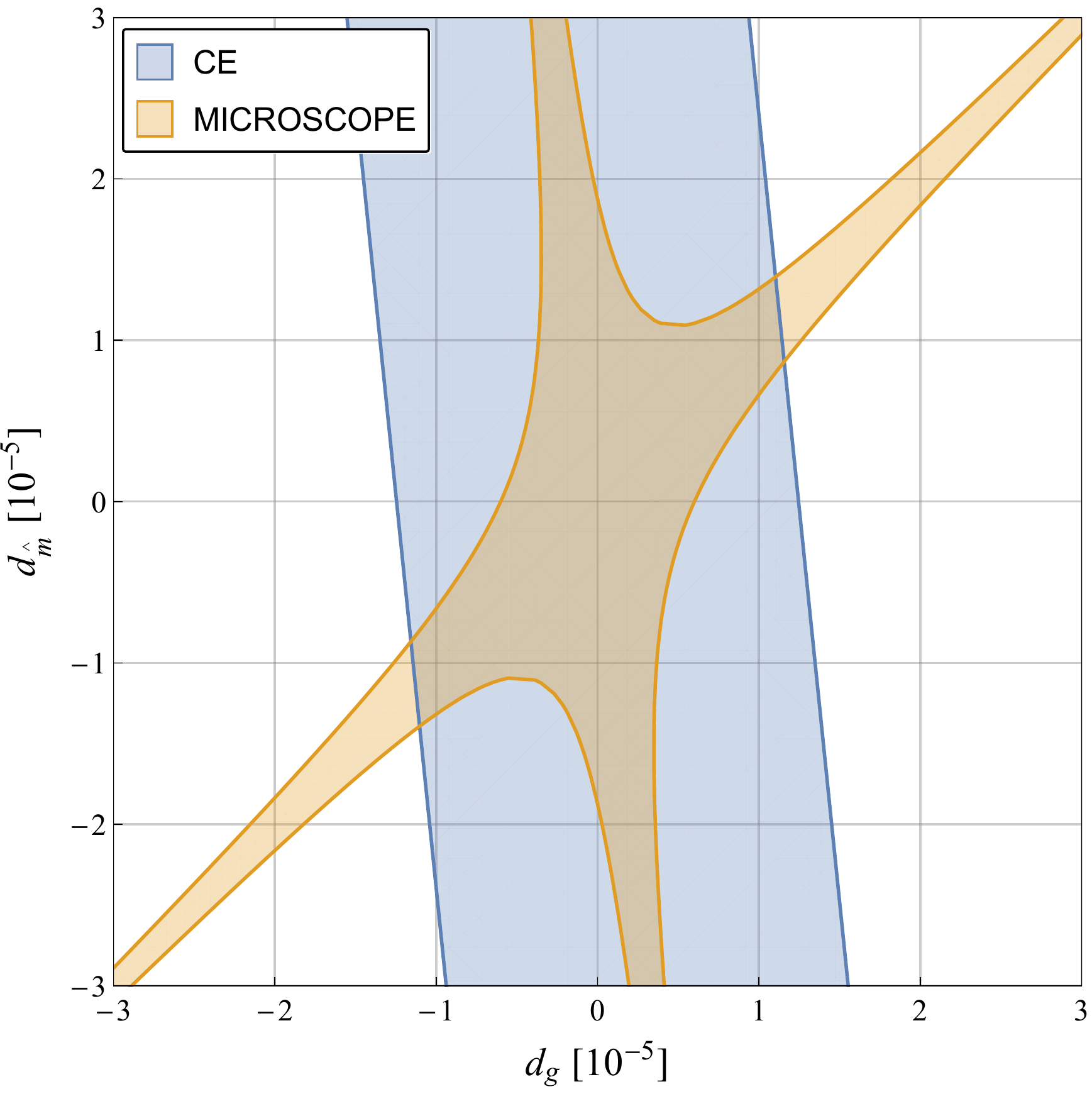}
        	\end{center}
      	\end{minipage}
      
    \end{tabular}
    \caption{The parameter regions which pass experiments with gravitational-wave detectors and the MICROSCOPE experiment are shown. The left figure shows the regions at $m_\phi = 10^{-16}~\mathrm{eV}$ and the right one shows those at $m_\phi = 10^{-13}~\mathrm{eV}$. We assume that $\Tobs = 1~\mathrm{year}$ and all of them are $2 \sigma$ limits.} 
    \label{EPcomparison}
  \end{center}
\end{figure}

On the other hand, the future constraints are weaker than those from the EP tests in most of the band.
However, it is the case only if the coupling constants other than $d_g$ are zero.
The EP tests constrain the following parameters,
\begin{equation}
D_{\hat{m}} = d^*_g (d_{\hat{m}}-d_g),~~~~~D_e = d^*_g d_e,
\end{equation}
while we can constrain $d^*_g$ with gravitational-wave detectors and fifth-force experiments.
Especially, EP tests give no constraints if $d_g = d_{m_e} = d_{m_u} = d_{m_d}$ and $d_e = 0$.
Fig. \ref{EPcomparison} shows the parameter regions which pass the MICROSCOPE experiment and experiments with gravitational-wave detectors in the case where $d_g$ and $d_{\hat{m}}$ are non-zero.
We can see that the latter experiments can exclude the parameter region along the line $d_g = d_{\hat{m}}$, which can not be excluded by MICROSCOPE.
In that sense, this search is complementary to the EP tests.

\section{Conclusion} \label{Conclusion}


We formulated a suitable data-analysis method to detect signals in outputs from gravitational-wave detectors and studied their detectability.
We first derived an analytical expression of the signal taking into account the motion of detectors and investigated its spectra.
Then, we formulated a suitable data-analysis method to detect this signal.
Taking into account the possibility that we have only one detector, we formulated a method which is applicable even with one detector.
Finally, we estimated the future constraints on the coupling constants between scalar field dark matter and Standard Model particles with our method.


As a result, we found that we can improve the previous estimate of the detectability given in \cite{Arvanitaki:2014faa} by a factor of $\sim 7$ at $f_\phi = 100~\mathrm{Hz}$.
We also found that all of the detectors we consider will provide more stringent constraints than the fifth-force experiments do in a mass band.
Especially, the constraints on $d_g$ at $m_\phi = 2 \times 10^{-17}~\mathrm{eV},~10^{-14}~\mathrm{eV}$ and $10^{-12}~\mathrm{eV}$ are improved by a factor of $\sim 30$, $\sim 100$ and $\sim 350$ respectively.
Since experiments with gravitational-wave detectors constrain a different combination of the coupling constants from that EP tests do, they play complementary roles.


Since we formulated a data-analysis method, the next thing to do is to apply the method to real data.
One of the issues to resolve for the application is line noise, which is sinusoidal noise and can mimic the signal we consider.
Such noises also degrade the sensitivity of continuous-wave searches \cite{Abbott:2018bwn}.
One characteristic of the signal to distinguish them can be expected bandwidth of $\sim f_\phi \vDM^2$.
Another issue is non-stationarity of the detectors caused by environmental disturbances such as earthquakes \cite{Biscans:2017yce}.
We need to generalize our method so that it can be applied to data, a part of which is unavailable due to increased noise level.
Improving our analysis method to resolve these issues and analyzing real data are left for the future works.

\acknowledgements{
We thank Kipp Cannon, Yousuke Itoh and Takashi Nakamura for helpful comments. 
This work was supported by a research program of the Advanced Leading Graduate Course for Photon Science (ALPS) at the University of Tokyo (S.M.), JSPS Grant-in-Aid for Young Scientists (B) No.15K17632 (T.S.) and MEXT Grant-in-Aid for Scientific Research on Innovative Areas No.15H05888 (T.S.), No.17H06359 (T.S.), and No.18H04338 (T.S.).
}

\appendix
\section{Metric perturbations generated by ultralight scalar field dark matter} \label{appendixGR}


In \cite{Aoki:2016kwl}, authors studied detectability of metric perturbations generated by ultralight scalar field dark matter.
We improve their estimate with respect to the following points.
\begin{itemize}
\item The previous study neglects sub-leading terms with respect to $m_\phi L$ ($h_1(t)$ in Eq. (\ref{GRNonLISA}) and Eq. (\ref{GRLISA})). However, since the leading term is suppressed by $\vDM \sim 10^{-3}$, the sub-leading terms are actually not negligible. We derive improved signal's formula by taking into account these sub-leading terms.
\item Since the signal is nearly monochromatic, we can improve signal-to-noise ratio by integrating the signal. We re-estimate the detectability with the analysis method we proposed in this paper.
\end{itemize}


Following calculations in \cite{Aoki:2016kwl}, we can easily find that a metric perturbation caused by monochromatic scalar waves, Eq. (\ref{Monotonic}), is
\begin{equation}
ds^2 = (1-2\Phi)(-dt^2 + d \vec{x}^2),
\end{equation}
where the oscillating part of $\Phi$ is
\begin{equation}
\Phi_{\mathrm{osc}}=\frac{\kappa^2 \rhoDM}{4 m^2_\phi} \mathrm{cos}\left[2 (\omegak t - \vec{k} \cdot \vec{x} + \thetak)\right].
\end{equation}
The motion of optical equipments caused by the metric perturbation is
\begin{equation}
\delta x^i (t, \vec{x}) \simeq \frac{\kappa^2 \rhoDM}{8 m^4_\phi} k^i \mathrm{sin}\left[2 ( \omegak t - \vec{k} \cdot \vec{x} + \thetak)\right].
\end{equation}
Following the calculations in Sec. \ref{response}, we can find that the signal caused by the metric perturbation is
\begin{align}
&h(t)=h_1(t)+h_2(t), \label{GRNonLISA} \\
&h_1(t) = \frac{\kappa^2 \rhoDM}{4 m^2_\phi} \frac{\mathrm{sin}^2(m_\phi L)}{m_\phi L} \frac{\vec{k} \cdot \vec{m} - \vec{k} \cdot \vec{n}}{m_\phi} \mathrm{sin} \left[2 \{ \omegak (t-L) - \vec{k} \cdot \vec{x} + \thetak \}\right],\\
&h_2(t) = -\frac{\kappa^2 \rhoDM}{4 m^2_\phi} \frac{(\veck \cdot \vec{m})^2 - (\veck \cdot \vec{n})^2}{m^2_\phi} \mathrm{cos}\left[2 \{ \omegak (t-L) - \veck \cdot \vec{x} + \thetak \}\right],
\end{align}
for detectors except for LISA and
\begin{align}
&h(t) = h_1(t) + h_2(t), \label{GRLISA} \\
&h_1(t) = -\frac{4 \kappa^2 \rhoDM}{m^2_\phi} \frac{\veck \cdot \vec{m} - \veck \cdot \vec{n}}{m_\phi} \mathrm{sin}^2(m_\phi L) \mathrm{sin}(2 m_\phi L) \mathrm{sin}(4 m_\phi L)\mathrm{cos}\left[ 2 \{ \omegak (t - 4 L) - \veck \cdot \vec{x} + \thetak \} \right], \\
&h_2(t) = -\frac{4 \kappa^2 \rhoDM L}{m_\phi} \frac{(\vec{k} \cdot \vec{m})^2 - (\veck \cdot \vec{n})^2}{m^2_\phi} \mathrm{sin}(2 m_\phi L) \mathrm{sin}(4 m_\phi L) \mathrm{sin}\left[ 2 \{ \omegak (t - 4 L) - \veck \cdot \vec{x} + \thetak \} \right],
\end{align}
for LISA.

\begin{figure}
	\begin{center}
		\includegraphics[width = 10cm]{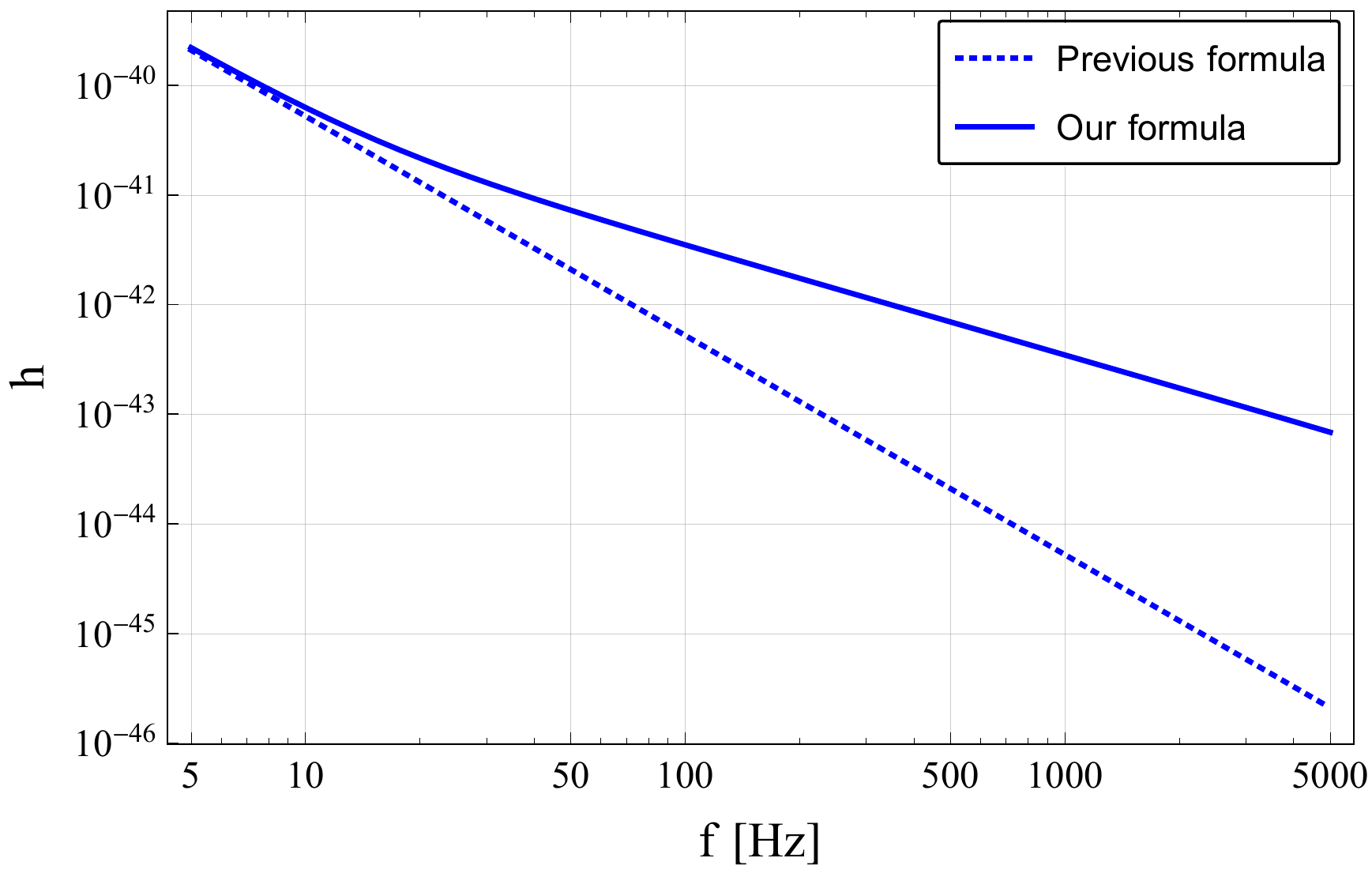}
		\caption{Amplitude of the signal generated by metric perturbations ultralight scalar field dark matter sources in LIGO data.}
	\label{ComparisonWithAoki}
	\end{center}
\end{figure}

While $h_1(t)$ is suppressed by $m_\phi L$ compared to $h_2(t)$ in the regime, $m_\phi L \ll 1$, $h_2(t)$ is also suppressed by $\vDM \sim 10^{-3}$ with respect to $h_1(t)$.
Therefore, over most of the frequency range we consider, $h_1(t)$ is more significant than $h_2(t)$.
Fig. \ref{ComparisonWithAoki} compares amplitudes of $h_2(t)$ and $h_1(t)+h_2(t)$ averaged over directions of $\vec{k}$ for LIGO.
We can see that the $h_1(t)$ becomes much larger as the signal's frequency increases.

\begin{figure}
	\begin{center}
		\includegraphics[width = 14cm]{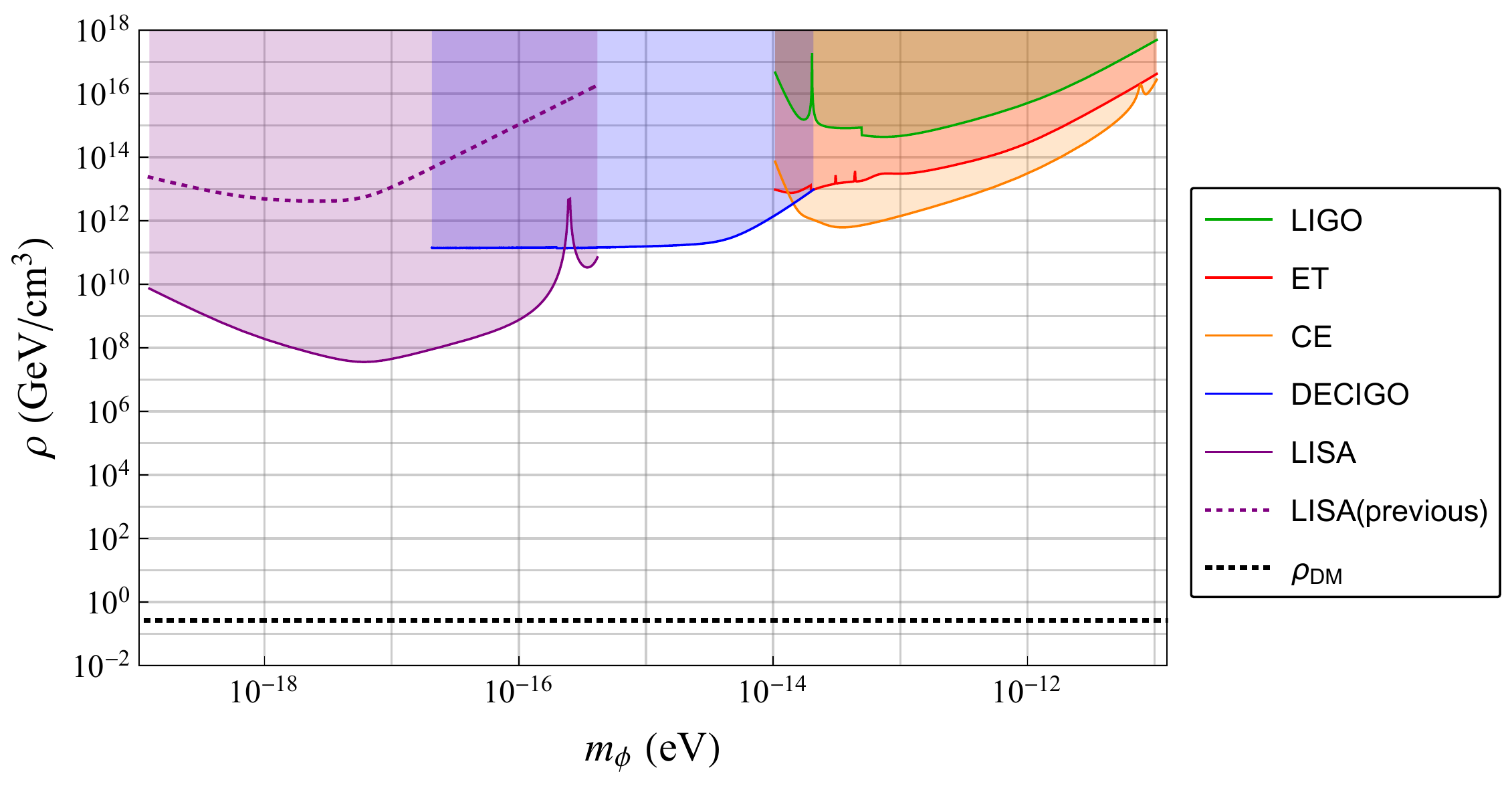}
		\caption{Constraints we will obtain with gravitational-wave detectors on the energy density of an ultralight scalar field which does not couple to Standard Model particles non-minimally.}
	\label{GRconstraint}
	\end{center}
\end{figure}

Finally, we re-estimate the detectability of the signal with our analysis method.
Fig. \ref{GRconstraint} shows the constraints on the energy density of an ultralight scalar field we can obtain with GW detectors.
It also shows the constraints by LISA obtained by the same method as that of the previous study (only
with $h_2(t)$ contributions and an integration time of $1$ second).
Although the constraints become much tighter, they are still far from $\rhoDM$.

\section{Modification of laser light propagation}\label{propagation}

In this section, we investigate how the scalar field affects the propagation of light through the coupling between the scalar field and photon.
The Lagrangian for photon is as follows,
\begin{equation}
\mathcal{L}_{\mathrm{photon}}= - \frac{1 + \psi}{4 e^2} F_{\mu \nu} F^{\mu \nu},~~~~\psi \equiv -d_e \kappa \phi.
\end{equation}
Then the equation of motion for the vector potential is 
\begin{equation}
\partial_\mu \left\{ (1 + \psi) \partial^\mu A^\nu \right\} - \partial_\mu \left\{ (1 + \psi) \partial^\nu A^\mu \right\} =0. \label{EOMphoton}
\end{equation}
In this paper, we consider the case where the scalar field behaves as the following non-relativistic matter waves in the Galaxy,
\begin{equation}
\psi = \psi_0 \cos (-m_\phi u^\nu x_\nu + \theta_0),~~~~u^\mu \simeq (1, ~\vec{v}),
\end{equation}
where $|\vec{v}| \ll 1$.
We obtain $A_{\mu}$ by solving Eq. (\ref{EOMphoton}) under the following assumption,
\begin{equation}
|\psi| \ll 1,~~~~~\frac{m_\phi}{2 \pi f_{\mathrm{laser}}} \ll 1. \label{approx_cond}
\end{equation}
The first condition means that the amplitude of the scalar field is tiny enough, which is valid in our case.
The second condition means the frequency of the laser light is much larger than the frequency of the oscillation we search for, which is valid in usual gravitational-wave experiments.

We separate the vector potential into two parts,
\begin{equation}
A^\mu = \bar{A}^\mu + \delta A^\mu,
\end{equation}
where the second part is the correction due to $\psi$.
To simplify the calculation, we apply a gauge transformation
\begin{equation}
A^\mu_{\mathrm{NEW}}=A^\mu_{\mathrm{OLD}}-\partial^\mu \Lambda.
\end{equation}
to move to a suitable gauge.
To obtain $\bar{A}^\mu$, we apply the Lorentz gauge,
\begin{equation}
\partial_\mu \bar{A}^\mu = 0.
\end{equation}
Then the equation of motion for $\bar{A}^\mu$ is 
\begin{equation}
\partial_\nu \partial^\nu \bar{A}^\mu = 0
\end{equation}
and its solution is
\begin{equation}
\bar{A}^\mu = a^\mu \mathrm{e}^{i k^\mu x_\mu},~~~~k^\mu k_\mu =0.
\end{equation}

To obtain $\delta A^\mu$, we apply a following gauge,
\begin{equation}
\partial_\mu \delta A^\mu + \partial_{\mu} \psi \bar{A}^\mu=0.
\end{equation}
To move to this gauge, we choose $\Lambda$ such that it satisfies
\begin{equation}
\partial_\mu \partial^\mu \Lambda = \partial_{\mu} \delta A^\mu_{\mathrm{OLD}} + \partial_\mu \psi \bar{A}^\mu. \label{gauge_cond}
\end{equation}
Since the right-hand side is the sum of two modes with four momenta of $k^\mu \pm m_\phi u^\mu$ and
\begin{equation}
(k^\mu \pm m_\phi u^\mu)(k_\mu \pm m_\phi u_\mu) \simeq \pm 4 \pi m_\phi f_{\mathrm{laser}} \neq 0,
\end{equation}
it is easy to show that Eq. (\ref{gauge_cond}) is solvable and
\begin{equation}
\Lambda = \left(\partial_\nu \partial^\nu\right)^{-1} \left[ \partial_\mu \delta A^\mu_{\mathrm{OLD}} + \partial_\mu \psi \bar{A}^\mu \right].
\end{equation}
In this gauge, the equation of motion for $\delta A^\mu$ is
\begin{equation}
\partial_\nu \partial^\nu \delta A^\mu + \partial_\nu \psi \partial^\nu \bar{A}^\mu + \partial_\nu \partial^\mu \psi \bar{A}^\nu = 0.
\end{equation}
Since the third term is negligible due to the second condition in Eq. (\ref{approx_cond}), this equation can be approximated as
\begin{equation}
\partial_\nu \partial^\nu \delta A^\mu + \partial_\nu \psi \partial^\nu A^\mu \simeq 0.
\end{equation}
The solution of this equation is
\begin{equation}
\delta A^\mu \simeq \frac{1}{2} \psi_0 a^\mu \cos \left[-m_\phi u^\mu x_\mu + \theta_0 \right] \mathrm{e}^{i k^\nu x_\nu}=-\frac{1}{2}d_e \kappa \phi a^\mu \mathrm{e}^{i k^\nu x_\nu}.
\end{equation}
Therefore, the solution for Eq. (\ref{EOMphoton}) is
\begin{equation}
A^\mu \simeq \left(1 - \frac{1}{2} d_e \kappa \phi(t,\vec{x}) \right)a^\mu \mathrm{e}^{i k^\nu x_\nu},
\end{equation}
which means the amplitude of the vector potential modulates.

Next we estimate the effect of this modulation on the output of the gravitational-wave detectors.
Due to the second condition in Eq. (\ref{gauge_cond}), the electromagnetic tensor can be approximated as
\begin{equation}
F_{\mu \nu} \simeq \left(1 - \frac{1}{2} d_e \kappa \phi(t,\vec{x}) \right)\bar{F}_{\mu \nu},~~~~~\bar{F}_{\mu \nu}\equiv \partial_\mu \bar{A}_\nu -\partial_\nu \bar{A}_\mu.
\end{equation}
Therefore laser power, $I$, modulates and the amplitude of this modulation is
\begin{equation}
\frac{\delta I}{I} \sim d_e \kappa \phi_{\mathrm{amp}}, 
\end{equation}
where $\phi_{\mathrm{amp}}$ is the amplitude of the scalar field's oscillation. 
Assuming the scalar field accounts for all the amount of dark matter and $\phi_{\mathrm{amp}} \sim \sqrt{2 \rhoDM}/m_\phi$, it can be estimated as
\begin{equation}
\frac{\delta I}{I} \sim 1 \times 10^{-16}d_e\left(\frac{1~\mathrm{Hz}}{f_\phi}\right),
\end{equation}
where $f_\phi=m_\phi/2 \pi$.
On the other hand, the power spectrum density of shot noise in $I$ is $S_I(f)=2 \hbar \omega I$, where $\omega$ is the angular frequency of laser light. 
Then 
\begin{equation}
\frac{\sqrt{S_I(f)}}{I} \simeq 2 \times 10^{-3}\left(\frac{1~\mathrm{W}}{I}\right)^{\frac{1}{2}} \left(\frac{100~\mathrm{nm}}{\lambda_{\mathrm{laser}}}\right)^{\frac{1}{2}}/\sqrt{\mathrm{Hz}},
\end{equation}
where $\lambda_{\mathrm{laser}}$ is the wavelength of laser light.
Since the time for which this signal keeps the coherence is $t_{\rm coh} \sim 1/f_\phi \vDM^2$, the detectable amplitude of this modulation is improved by a factor of
$1/\sqrt{t_{\rm coh}} \sim (f_\phi \vDM^2)^{\frac{1}{2}}$,
\begin{equation}
\frac{\delta I_{\mathrm{th}}}{I}\sim \frac{1}{\sqrt{t_{\rm coh}}}\frac{\sqrt{S_I(f)}}{I}
\sim
2 \times 10^{-6} \left(\frac{f_\phi}{1~\mathrm{Hz}}\right)^{\frac{1}{2}} \left(\frac{1~\mathrm{W}}{I}\right)^{\frac{1}{2}}\left(\frac{100~\mathrm{nm}}{\lambda_{\mathrm{laser}}}\right)^{\frac{1}{2}}.
\end{equation}
Therefore, for the modulation due to the scalar field's oscillation to be detected, the laser power must satisfy
\begin{equation}
I \gtrsim \frac{4 \times 10^{19}}{d^2_e} \left(\frac{1000~\mathrm{nm}}{\lambda_{\mathrm{laser}}}\right)\left(\frac{f_\phi}{1~\mathrm{Hz}}\right)^3~\mathrm{W}. \label{lasercondition}
\end{equation}
In the experiments we consider, $\lambda_{\mathrm{laser}} \lesssim \mathcal{O}(1000)~\mathrm{nm}$.
On the other hand, the Equivalence Principle tests \cite{Schlamminger:2007ht,Wagner:2012ui,Touboul:2017grn} have already provided the constraints, $d_e \lesssim 10^{-2}$ for the ground-based detectors' band and $d_e \lesssim 10^{-4}$ for the space-based detectors' band, if it is assumed that miraculous cancellation between the coupling constants does not occur.
Therefore, the required laser power (\ref{lasercondition}) is much higher than laser power in the gravitational-wave experiments we consider.
In reality, we can not integrate the signal for longer time than the observational time. 
It makes the requirement for laser power more severe. 
On the other hand, if $1/f_\phi \vDM^2$ is shorter than observational time, we can improve S/N with the analysis method we proposed by a factor of $N^{\frac{1}{4}} (f_\phi)$ as mentioned below Eq. (\ref{threshold_incoherent2}).
However, the difference is $\mathcal{O}(1)$ and it does not change the conclusion.
Therefore, the modification of laser light propagation does not cause detectable signals.

\section{Time variation of the detectors' orientations} \label{timevariation}
\begin{figure}
  \begin{center}
    \begin{tabular}{c}

      \begin{minipage}{0.5\hsize}
        \begin{center}
          \includegraphics[clip, width=8cm]{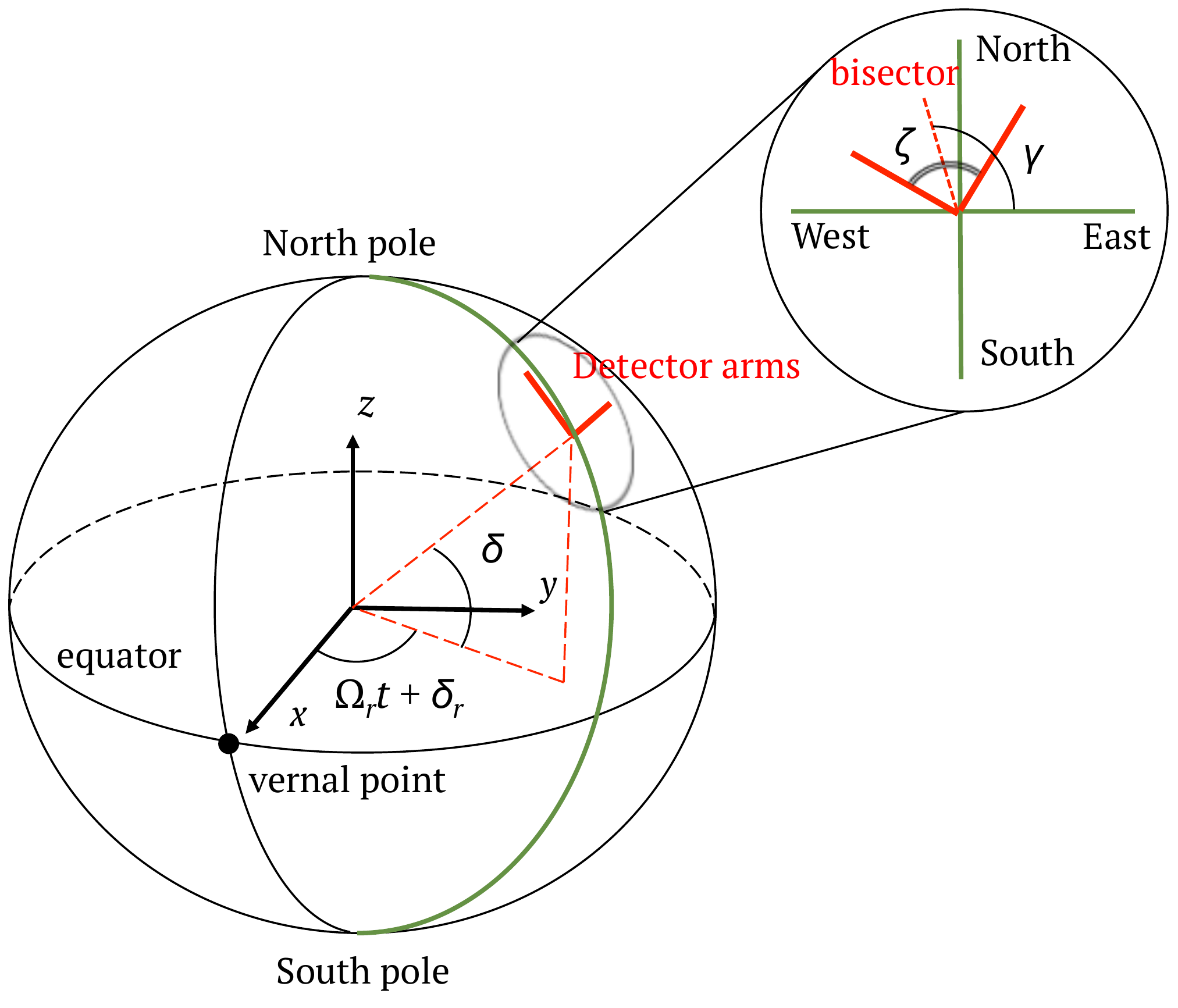}
        \end{center}
      \end{minipage} 

      \begin{minipage}{0.5\hsize}
 	      \begin{center}
    	     \includegraphics[clip, width=8cm]{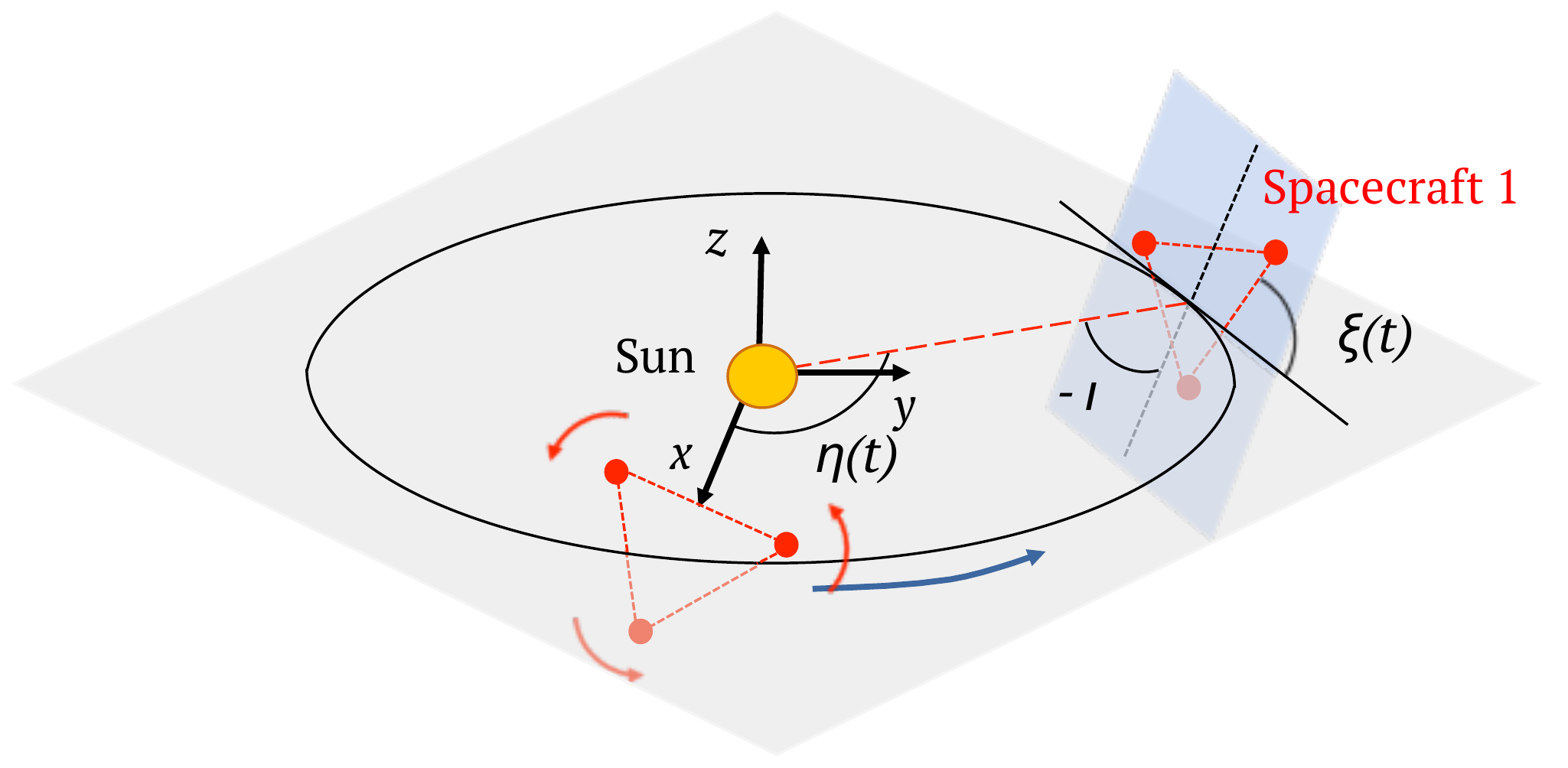}
        	\end{center}
      	\end{minipage}
      
    \end{tabular}
    \caption{The configurations of a ground-based detector (Left) and a space-based detector (Right) are shown.} 
    \label{DetectorConfigration}
  \end{center}
\end{figure}

In this appendix, we obtain the expressions of the functions in the right-hand sides of Eq. (\ref{hhg}), (\ref{hlg}), (\ref{hhd}), (\ref{hld}), (\ref{Xh}) and (\ref{Xl}).
Here we refer to Appendix C of \cite{Jaranowski:2009zz}.
The detectors' configurations are shown in Fig. \ref{DetectorConfigration}.

For ground-based detectors, we consider the coordinate, $(t,\vec{x})$, as a celestial coordinate whose origin is the center of the Earth, whose z axis coincides with the Earth's rotation axis and points toward the North pole, whose x and y axes lie in the Earth's equatorial plane, and whose x axis points toward the vernal point.
Then, we have
\begin{equation}
\vec{n}(t)=O^{\mathrm{T}}_2(t) O^{\mathrm{T}}_3 \hat{\vec{n}},~~~~\vec{m}(t)=O^{\mathrm{T}}_2(t) O^{\mathrm{T}}_3 \hat{\vec{m}}.
\end{equation}
The matrices and vectors in the right-hand sides are
\begin{equation}
O_2(t)=
\left(
\begin{array}{ccc}
\sin \delta \cos(\deltar+\Od t) & \sin \delta \sin(\deltar + \Od t)  & -\cos \delta  \\
 -\sin(\deltar + \Od t)& \cos(\deltar + \Od t) & 0\\
\cos \delta \cos(\deltar + \Od t) &   \cos \delta \sin(\deltar + \Od t) & \sin \delta
\end{array}
\right),
\end{equation}
\begin{equation}
O_3=
\left(
\begin{array}{ccc}
-\sin\left(\gamma - \frac{\zeta}{2}\right) & \cos\left(\gamma-\frac{\zeta}{2}\right)  & 0  \\
-\cos\left(\gamma - \frac{\zeta}{2}\right)& -\sin\left(\gamma-\frac{\zeta}{2}\right) & 0\\
0 & 0 & 1
\end{array}
\right),
\end{equation}
and
\begin{equation}
\hat{\vec{n}}=(1,0,0)^{\mathrm{T}},~~~~\hat{\vec{m}}=(\cos \zeta, \sin \zeta, 0)^{\mathrm{T}},
\end{equation}
where $\delta$ is the geodetic latitude of the detector's site, $\deltar$ is the phase defining the position of the Earth in its diurnal motion at $t=0$, $\gamma$ is the angle measured counter-clockwise from East to the bisector of the interferometer arms and $\zeta$ is the angle between the interferometer arms.
Then the functions in the right-hand sides of Eq.  (\ref{hhg}) and (\ref{hlg}) are
\begin{align}
g_{1,0}(t)&=-2  \cos \gamma \cos \delta \sin\left(\frac{\zeta}{2}\right) \partial_z \phi,\\
g_{1,1}(t)&=\e^{i \deltar}(\sin \gamma - i \cos \gamma \sin \delta ) \sin\left(\frac{\zeta}{2}\right)(i \partial_x + \partial_y) \phi, \\
g_{2,0}(t)&= \cos \gamma \sin \gamma \cos^2 \delta \sin \zeta (\partial_{xx} + \partial_{yy}- 2 \partial_{zz} ) \phi,  \\
g_{2,1}(t)&=-\e^{i \deltar} \cos \delta (\cos(2 \gamma)+i\sin(2 \gamma) \sin \delta ) \sin \zeta (i \partial_{xz} +\partial_{yz}) \phi, \\
g_{2,2}(t)&=\frac{1}{2} \e^{2 i \deltar} (i \sin \gamma + \cos \gamma \sin \delta)(\cos \gamma + i \sin \gamma \sin \delta ) \sin \zeta (i \partial_{xx} + 2 \partial_{xy} - i \partial_{yy})\phi,
\end{align}
where $\partial_{i_1i_2\dots i_n} \equiv \partial_{i_1} \partial_{i_2} \dots \partial_{i_n}$.

For space-based detectors, we consider the coordinate, $(t,\vec{x})$, as an ecliptic coordinate whose origin is the center of the Sun.
Then we have
\begin{equation}
\vec{n}(t)=O_2(t) \hat{\vec{n}},~~~~\vec{m}(t)=O_2(t) \hat{\vec{m}}.
\end{equation}
The matrix and vectors in the right-hand sides are
\begin{equation}
O_2(t)=
\left(
\begin{array}{ccc}
\sin \eta(t) \cos \xi(t) - \sin \iota \cos \eta(t) \sin \xi(t) & -\sin \eta(t) \sin \xi(t) - \sin \iota \cos \eta(t) \cos \xi(t)  & -\cos \iota \cos \eta(t) \\
-\cos \eta(t) \cos \xi(t) - \sin \iota \sin \eta(t) \sin \xi(t) & \cos \eta(t) \sin \xi(t) - \sin \iota \sin \eta(t) \cos \xi(t) & -\cos \iota \sin \eta(t) \\
\cos \iota \sin \xi(t) &  \cos \iota \cos \xi(t) & -\sin \iota
\end{array}
\right),
\end{equation}
and
\begin{equation}
\hat{\vec{n}}=-(\cos \sigma_3, \sin \sigma_3, 0),~~~~\hat{\vec{m}}=(\cos \sigma_2, \sin \sigma_2,0),
\end{equation}
where
\begin{equation}
\iota=-\frac{\pi}{6},~~~~\eta(t)=\Oy t + \eta_0,~~~~\xi(t)=-\Oy t + \xi_0,
\end{equation}
and
\begin{equation}
\sigma_a = -\frac{3}{2}\pi+\frac{2}{3}(a-1)\pi.
\end{equation}
Then the functions in the right-hand sides of Eq. (\ref{hhd}), (\ref{hld}), (\ref{Xh}) and (\ref{Xl}) are
\begin{align}
c_{1,0}(t)&=\frac{3}{4} ( \cos(\eta_0+\xi_0) \partial_x \phi + \sin(\eta_0 + \xi_0) \partial_y \phi) \\
c_{1,1}(t)&= \frac{\sqrt{3}}{4} \e^{-i \xi_0} \partial_z \phi \\
c_{1,2}(t)&=-\frac{1}{8} \e^{i(\eta_0 - \xi_0)} (\partial_x - i \partial_y) \phi,\\
c_{2,0}(t)&=\frac{9 \sqrt{3}}{32} (2 \cos(2 (\eta_0 + \xi_0)) \partial_{xy} \phi + \sin(2 (\eta_0 + \xi_0)) (- \partial_{xx}+\partial_{yy}) \phi)  \\
c_{2,1}(t)&=\frac{9}{16} \e^{-i(\eta_0 + 2 \xi_0)} ( -i \partial_{xz} + \partial_{yz}) \phi \\
c_{2,2}(t)&= \frac{3 \sqrt{3}}{32} i \e^{-2 i \xi_0} (\partial_{xx} + \partial_{yy} - 2 \partial_{zz}) \phi  \\
c_{2,3}(t)&= \frac{3}{16} \e^{i(\eta_0 - 2 \xi_0)} ( i \partial_{xz} + \partial_{yz}) \phi \\
c_{2,4}(t)&= - \frac{\sqrt{3}}{64} i \e^{2 i (\eta_0 - \xi_0)}(\partial_{xx}-2 i \partial_{xy} - \partial_{yy})\phi,\\
x_{1,0}(t)&=\frac{3}{4} ( \cos(\eta_0+\xi_0) \partial_{tx} \phi + \sin(\eta_0 + \xi_0) \partial_{ty} \phi) \\
x_{1,1}(t)&= \frac{\sqrt{3}}{4} \e^{-i \xi_0} \partial_{tz} \phi \\
x_{1,2}(t)&=-\frac{1}{8} \e^{i(\eta_0 - \xi_0)} (\partial_{tx} - i \partial_{ty}) \phi,\\
x_{2,0}(t)&=\frac{9 \sqrt{3}}{32} (2 \cos(2 (\eta_0 + \xi_0)) \partial_{txy} \phi + \sin(2 (\eta_0 + \xi_0)) (- \partial_{txx}+\partial_{tyy}) \phi)  \\
x_{2,1}(t)&=\frac{9}{16} \e^{-i(\eta_0 + 2 \xi_0)} ( -i \partial_{txz} + \partial_{tyz}) \phi \\
x_{2,2}(t)&= \frac{3 \sqrt{3}}{32} i \e^{-2 i \xi_0} (\partial_{txx} + \partial_{tyy} - 2 \partial_{tzz}) \phi  \\
x_{2,3}(t)&= \frac{3}{16} \e^{i(\eta_0 - 2 \xi_0)} ( i \partial_{txz} + \partial_{tyz}) \phi \\
x_{2,4}(t)&= - \frac{\sqrt{3}}{64} i \e^{2 i (\eta_0 - \xi_0)}(\partial_{txx}-2 i \partial_{txy} - \partial_{tyy})\phi.
\end{align}

\bibliographystyle{unsrt}
\bibliography{UltralightScalarFieldDM}

\end{document}